\documentclass[aps,twocolumn,pre,showpacs,nofootinbib,superscriptaddress,nofootinbib]{revtex4}
\usepackage{graphicx}

\newcommand{\MCtwo}{Microtechnology and Nanoscience, MC2, 
Chalmers University of Technology, SE-412 96 G{\"o}teborg, Sweden}
\newcommand{\chalmersTF}{Department of Applied Physics,
Chalmers University of Technology,
SE-41296 G\"{o}teborg, Sweden}
\newcommand{\GUfysik}{Department of Physics, G\"oteborg University, SE-41296 G\"oteborg, Sweden}
\newcommand{\rutgers}{Department of Physics and Astronomy, Rutgers University,
Piscataway, NJ 08854-8019, USA}
\newcommand{\oak}{Materials Science and Technology Division, Oak Ridge National
Laboratory, Oak Ridge, Tennessee 37831, USA}

\newcommand{\vdW}{{\mbox{\scriptsize vdW}}}
\newcommand{\myatop}{{\mbox{\scriptsize atop}}}

\newcommand{\mymin}{{\mbox{\scriptsize min}}}
\newcommand{\myhom}{{\mbox{\scriptsize hom}}}
\newcommand{\hollow}{{\mbox{\scriptsize hollow}}}
\newcommand{\bridgea}{{\mbox{\scriptsize long-bridge}}}
\newcommand{\nl}{{\mbox{\scriptsize nl}}}
\newcommand{\bridgeb}{{\mbox{\scriptsize short-bridge}}}
\newcommand{\He}{{\mbox{\scriptsize He}}}
\newcommand{\vdWDFtwo}{{\mbox{\scriptsize vdW-DF2}}}
\newcommand{\EXX}{{\mbox{\scriptsize EXX}}}

\begin{document}

\title{Benchmarking van der Waals Density Functionals with Experimental 
Data:\\
Potential Energy Curves for H$_2$ Molecules on Cu(111), (100), and 
(110) Surfaces}

\author{Kyuho Lee}\affiliation{\rutgers}
\author{Kristian Berland}\affiliation{\MCtwo}
\author{Mina Yoon}\affiliation{\oak}
\author{Stig Andersson}\affiliation{\GUfysik}
\author{Elsebeth Schr{\"o}der}\thanks{Corresponding author}%
\email{schroder@chalmers.se}%
\affiliation{\MCtwo}
\author{Per Hyldgaard}\affiliation{\MCtwo}
\author{Bengt I. Lundqvist}\affiliation{\chalmersTF}

\date{May 16, 2012}

\begin{abstract}
Detailed physisorption data from experiment for the H$_2$ 
molecule on low-index Cu surfaces challenge theory.
Recently, density-functional theory (DFT) has been developed to account 
for nonlocal correlation effects, including van der Waals (dispersion) 
forces. We show that the functional vdW-DF2 gives a potential-energy 
curve, potential-well energy levels, and difference in lateral corrugation promisingly close 
to the results obtained by resonant elastic backscattering-diffraction 
experiments. The backscattering barrier is found selective 
for choice of exchange-functional approximation. 
Further, the DFT-D3 and TS-vdW corrections to traditional DFT formulations 
are also benchmarked, and deviations are analyzed.
\end{abstract}

\maketitle


\section{Introduction} 

The van der Waals (vdW) or dispersion interactions play 
important roles in defining structure, stability, and function for molecules 
and materials. 
Our understanding of chemistry, biology, solid state physics, and materials 
science benefits greatly from density-functional theory (DFT). 
This in principle exact theory for stability and structure of electron 
systems \cite{Kohn} is computationally feasible also for 
complex and extended systems. However, in practice, approximations have 
to be made to describe exchange and correlation (XC) of the participating 
electrons \cite{KohnSham}. This work aims at benchmarking some XC descriptions 
of nonlocal correlations that describe vdW interactions 
\cite{LeeEtAl10,Grimme,TS}. 

To boil down the intricate electronic dynamics behind the vdW interaction 
into a density functional $F[n]$ is a formidable task. $F[n]$ should 
depend only on the electron density $n(\mathbf{r})$ and do that in the right and 
generally applicable way. It should obey fundamental physical laws, 
like charge conservation and time invariance, and have a physically 
sound account of system and interactions. The vdW-DF 
method \cite{dion2004,LeeEtAl10} has such ambitions. There are more pragmatic 
methods, including those correcting traditional DFT calculations with pairwise 
vdW potentials, like the DFT-D \cite{Grimme} and TS-vdW \cite{TS} methods. 
Further, first-principles electron-structure calculations
are made efficient but still carry
much higher computational costs than DFT.
An example is
the random-phase approximation (RPA) to the correlation energy used as a 
suitable complement to the exact exchange energy \cite{HarlKresse2009}. 

The results obtained from particular XC functionals and other vdW descriptions
can be assessed by comparing with other accurate electron-structure theories
like those presented in Refs.\ \onlinecite{Jurecka2006,Takatani2010}, 
or with experiments. 
Typically one or two measurable quantities are available, like in 
Ref.\ \onlinecite{Svetla}.
In Ref.\ \onlinecite{LeeEtAl10} some of us stressed the importance of 
exploiting
extended and accurate experimental data sets when these are available.
Here, we extend this comparison by considering several facets of the Cu
surface. 

Surface physics has a long and successful tradition of detailed and 
informative experiment-theory comparisons and offers possibilities also here. 
Extensive data sets are available for systems and conditions where the 
weak vdW forces can be reached and accurately mapped. 
A full physisorption potential and a detailed characterization thereof 
have been derived from versatile, accurate, and clearly interpretable 
measurements. In the physisorption regime, resonant elastic 
backscattering-diffraction experiments from low-index crystal faces 
provide a detailed quantitative knowledge. 
The actual data bank is rich and covers results, for instance for
the whole shape of the physisorption potential, for
the differences in corrugation across several facets, 
and for the energy levels in the potential well.

This Paper compares state-of-the-art vdW 
descriptions of physisorption of H$_2$ and D$_2$ molecules on 
the low-indexed Cu surfaces with  
physisorption potentials constructed from selective-adsorption 
bound-state measurements.
These data were analyzed in the early 90s 
in model systems \cite{andersson1993,perandersson1993,persson2008}. 
In general terms, the measurements, calculations, and analysis
underline the importance of building in the 
essential surface-physics into vdW functionals and other vdW accounts. 

An earlier study of H$_2$ on the close-packed Cu(111) surface shows 
some spread in the results from different vdW accounts and that 
one of the tested XC functionals (vdW-DF2) compares 
promisingly with the experimental physisorption potential \cite{h2cu}. 
This motivates an extension of the study to the hydrogen molecule on other, 
more corrugated Cu surfaces. 
This paper is a significant extension of Ref.\ \onlinecite{h2cu}, addressing 
questions that were not resolved back in 1993 
\cite{andersson1993,perandersson1993,persson2008}, for instance, 
trends with crystal face. 

The outline, beyond this introduction, is as follows: First a brief 
review of physisorption, in particular on metal surfaces, and a review 
of the traditional description. This is followed by a presentation of 
some DFTs with accounts of vdW forces, a presentation of the systems 
studied, and a review of the experimental benchmark sets. Next,
calculated results for Potential-Energy Curves (PECs) and other 
physical quantities are presented, and the Paper is concluded with 
comparisons, analysis, and outlook for future functionals.


\section{Physisorption and Weak Adsorption\label{sec:2}} 

Chemically inert atoms and molecules adsorb physically on cold metal 
surfaces \cite{persson2008}. Characteristic desorption temperatures 
range from only a few K to tens of K, while adsorption energies range from a few meV 
to around 100 meV. These values may be determined from measurements of thermal 
desorption and isosteric heat of adsorption. For light adsorbates, like He and H$_2$,  
gas-surface-scattering experiments provide 
a more direct and elegant method which involves the elastic 
backscattering with resonance structure. The bound-level sequences in 
the potential well can be measured with accuracy and in detail. 
Isotopes with widely different masses ($^3$He, $^4$He, H$_2$, D$_2$) 
are available. This permits a unique assignment of the levels and a 
determination of the well depth and ultimately a qualified test of model 
potentials~\cite{Roy}. 

The potential well is formed by the vdW attraction 
which arises from adsorbate-substrate electron correlation.
At large distances 
from the surface the vdW attraction goes like $V_\vdW(z) = -C_\vdW/z^3$. 
Here $z$ is the distance normal to the surface, measured from the 
center-of-mass of the particle to a surface reference plane
close to the outermost layer of ion-cores in the solid, the so-called vdW plane.
Near the surface the short-range repulsion, the ``corrugated wall", acts.

Specifically, we consider molecules that physisorb on metal surfaces 
where no significant change in the electronic configuration takes place upon 
adsorption. The weak coupling to electronic excitations 
\cite{SchGunnars1980} makes the adsorption largely electronically 
adiabatic. The energy transfer occurs through the phonon system of the 
solid lattice \cite{Brenig1987}. These conditions are expected to hold 
for hydrogen molecules on simple or noble metals. 

In early days, atom- and molecule-diffraction studies of metallic 
single-crystal surfaces were lagging behind those of ion-crystal surfaces. 
On metals, diffraction spots appear much weaker, which reflects the much 
weaker corrugation of close-packed metal surfaces \cite{Boato}, than on 
an ionic crystal, like LiF(100) \cite{Garcia}.

In the traditional picture of physisorption, the interaction between an 
inert adparticle and a metal surface is approximated as 
a superposition of the long-ranged $V_\vdW$ 
and a short-range Pauli repulsion, $V_R$.
The latter is due to the overlap between 
wavefunction tails of the metal conduction electrons and the 
closed-shell electrons of the adparticle 
\cite{zaremba1977,HarrisNordlander,persson2008},
\begin{equation}
V_0(z) = V_R (z) + V_\vdW(z).
\label{eq:1}
\end{equation}
Here approximately
\begin{equation}
V_R (z) = V^\prime_R \exp(-\alpha z),  
\label{eq:2}
\end{equation}
and 
\begin{equation}
V_\vdW(z) = - \frac{C_\vdW }{(z - z_\vdW)^3} f(2k_c(z - z_\vdW)),
\label{eq:3}	
\end{equation}
now with $z$ measured from the ``jellium" edge \cite{jelliumedge}.
$V_0(z)$ is an effective potential. It arises as a lateral and 
adsorption-angle average of an underlying adsorption potential.  
We use  $V_1(z)$ to express the amplitude of the modulation around the 
average $V_0(z)$. 

The repulsive potential $V_R(z)$ has a prefactor $V^\prime_R$ that can be 
determined from the shifts of the metal one-electron energies caused by 
the adparticle. It can also be calculated by, for example perturbation theory in a 
pseudo-potential description of the adparticle and a jellium-model 
representation of the metal surface \cite{andersson1993}.  

The strength of the asymptotic vdW attraction, $C_\vdW$, and 
the reference-plane position, $z_\vdW$, depend on the dielectric 
properties of the metal substrate and the adsorbate \cite{ZarembaKohn1976,Liebsch1986}. 
The prefactor $f(z)$ in the potential $V_\vdW(z)$ of Eq.\ (\ref{eq:3}) 
introduces a saturation of the attraction at atomic-scale 
separations. The function $f(x)$ [$f(x)=1-(1+x+x^2/2)\exp(-x)$ in some accounts] 
lacks a rigorous prescription and thus includes some level of arbitrariness 
for $V_\vdW(z)$. Experimental data provides a possible empirical 
solution to this dilemma. 

The physisorption potential $V_0(z)$ in Eq.\ (\ref{eq:1}) 
depends on the details of the surface electron structure 
both via the electron spill out ($V_R$) and the 
spatial decay of polarization properties in the surface region ($V_\vdW$). 
Accordingly, there is a crystal-face dependence of $V_0(z)$ for a given 
adparticle \cite{persson2008}. 

\begin{figure*}
\begin{center}
\includegraphics[width=0.85\textwidth]{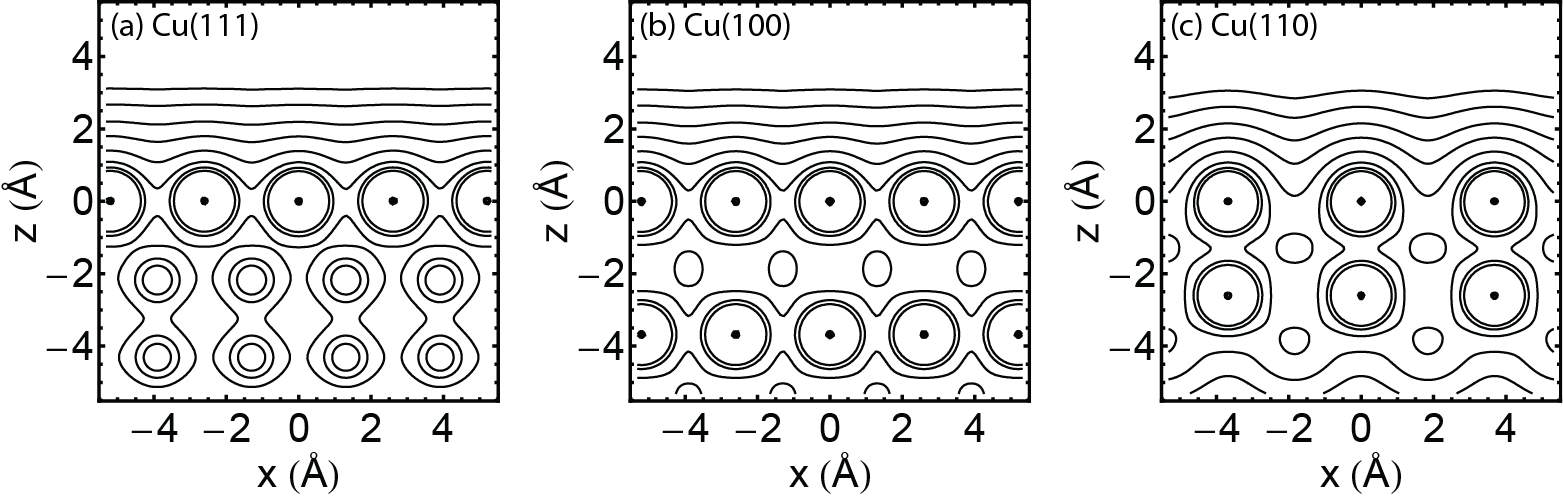}
\caption{\label{fig:3}
Electron density profiles $n_o(\mathbf{r})$ of the clean Cu(111), (100), and
(110) surfaces calculated with the vdW-DF2 functional.
The density contours take values in a  nonlinear fashion.
}
\end{center}
\end{figure*}

Figure \ref{fig:3} shows the electron density profiles calculated for the Cu(111),
(100), and (110) surfaces with the method described below.
They illustrate that the corrugations on these facets are small but
differ, growing in order (111) $<$ (100) $<$ (110).
For the scattering experiment, the density far out in the tails is
particularly important.

That He-atom diffraction from dense metal surfaces is weak (compared 
to for instance ionic crystals) was early observed \cite{Boato} and 
subsequently 
explained in terms of a simple tie between the scattering potential 
and the electron-density profile: The He-surface interaction energy 
$E_\He(r)$ can reasonably well be expressed as \cite{Esbjerg}
\begin{equation}
E_\He(\mathbf{r}) \simeq E_\He^\myhom(n_o(\mathbf{r})),
\label{eq:4}
\end{equation}
where $E_\He^\myhom(n)$ is the energy change on embedding a free He atom 
in a homogeneous electron gas of density $n$, and $n_o(\mathbf{r})$ is the host 
electron density at point $\mathbf{r}$. On close-packed metal surfaces the 
electron distribution $n_o(\mathbf{r})$ is smeared out almost uniformly along 
the surface \cite{Smoluchowski1941}, thus giving weak corrugation. 
The crude proposal (\ref{eq:4}) might be viewed as the precursor to 
the effective-medium theory \cite{EMT}. 	

The form (\ref{eq:4}) provides an interpretation of the mechanism 
by which the increasing density corrugation for (111) $<$ (100) $<$ (110)
causes increasing amplitudes of modulation $V_1(z)$ in the 
physisorption potentials. 
The min-to-max variation
of the rotationally averaged, lateral periodic corrugation
$V_1(z)$  is modeled with an amplitude function \cite{HarrisAndLiebsch1982b},
like in Eq.\ (\ref{eq:2}), $V_1(z) = V_1^\prime \exp(-\beta z)$. Here the
exponent $\beta$ is related to the exponent $\alpha$ of $V_0(z)$
via $\beta = \alpha/2 + \sqrt{(\alpha/2)^2 + G_{10}^2}$.
The strength prefactor $V_1^\prime$ is adjusted so that the calculated 
intensities of the first-order $\mathbf{G}_{10}$ diffraction beams 
agree with measured values.

\begin{figure}
\begin{center}
\includegraphics[width=0.4\textwidth]{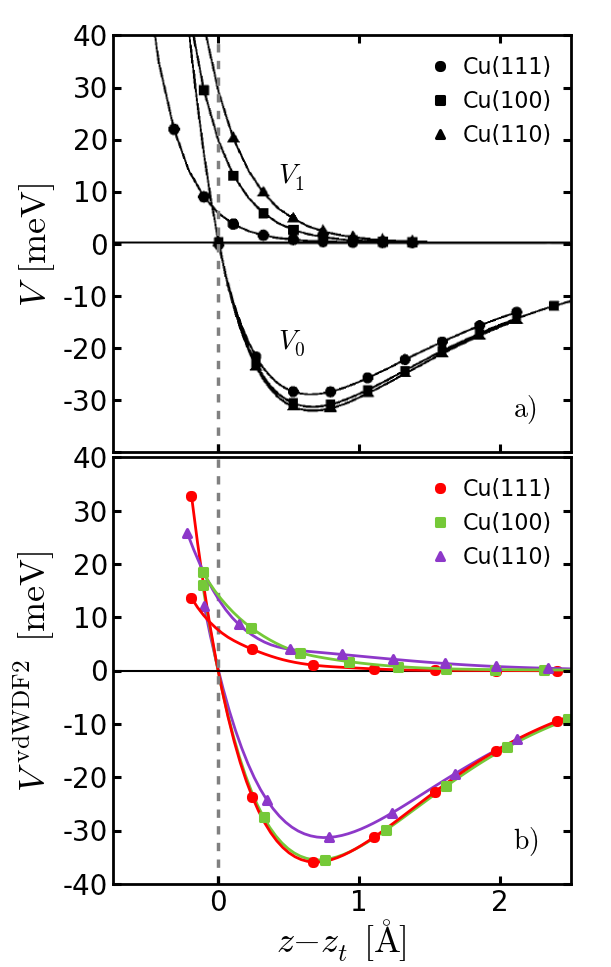}
\caption{\label{fig:1}
(a) Physisorption interaction potentials, $V$, for H$_2$ (D$_2$) on Cu(111)
(circles), Cu(100) (squares), and Cu(110) (triangles), in the form of
lateral average, $V_0(z)$ and corrugation $V_1(z)$.
The potential functions $V_0$ and $V_1$ are defined
in the text. The position $z$ of the molecular center of mass is here
given with respect to the classical turning point, i.e. the position
$z_t$ where a classical particle at energy $\epsilon_i = 0$ would be 
reflected in the potential.
Adapted from \cite{andersson1993}.
(b) The corresponding PECs
calculated with vdW-DF2, averaged according
to Eq.\ (\protect\ref{eq:7}).}
\end{center}
\end{figure}

The simple message of the experimental characterization in
Figure \ref{fig:1}(a) is that, at the optimal separation and
out, the $V_1$ corrugation terms are rather weak
compared to the $V_0$ averages. This observation confirms that the basic
particle-surface interaction is predominantly one dimensional.

\begin{figure}
\begin{center}
\includegraphics[width=0.4\textwidth]{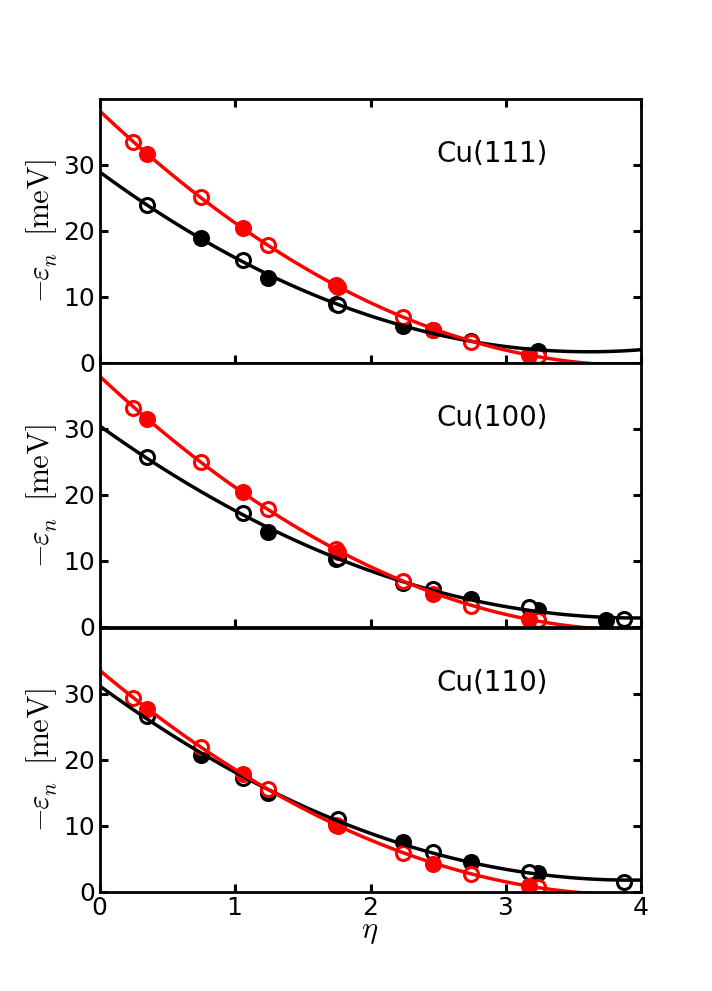}
\caption{\label{fig:2}
Energy levels in the $V_0$ potentials of Fig.~\ref{fig:1}(b) are plotted
versus the mass-reduced level number $\eta= (n + 1/2)/\sqrt{m}$,
where $n$ is the quantum number and $m$ the mass.  Experimentally
determined values are given by solid black and open black circles (the open
circles being deuterium results).
The theory results for the levels are
identified by solid red and open red circles (the open circles being the
deuterium results). Experimental curves adapted from \cite{andersson1993}.
}
\end{center}
\end{figure}

Figure \ref{fig:1}(a) shows experiment-based PECs $V_0(z)$ for physisorption of H$_2$ on 
the Cu(111), Cu(100), and Cu(110) surfaces. The H$_2$ molecules are trapped 
in states $\epsilon_i$ that are quantized in the perpendicular direction but have 
an essentially free in-surface dynamics. The panel details the laterally (and 
rotationally) averaged potential $V_0(z)$ that reflects the perpendicular quantization,
i.e., the physisorption levels $\epsilon_n$. The experiment-based forms of $V_0(z)$ [Fig.~\ref{fig:1}(a)] 
are obtained by adjusting the parameters\footnote{The procedure takes off from the Le Roy 
analysis \cite{Roy,andersson1993} that gives an approximate determination of $C_\vdW$  
(a complete direct specification of $C_\vdW$ would require measurements of even more 
shallow quantized physisorption levels) and of the depth of the physisorption 
well \cite{perandersson1993}. It also takes off from an approximate Zaremba-Kohn 
type \cite{zaremba1977,HarrisNordlander} specification of the repulsive wall.  The location of 
the jellium edge \cite{jelliumedge} (relative to the position of the last atomic plane) is set at 
1.97 $a_0$, at 1.71 $a_0$, and at 1.21 $a_0$ for the Cu(111), Cu(100) and for Cu(110) surfaces.  
The remaining set of parameters are split into two groups, those that are constrained to 
be identical for all facets and those that are assumed to be facet specific. Fitting against 
the set of measured quantization levels $\epsilon_n$ yields values $C_\vdW=4740 a_0^3$ meV, 
$z_\vdW= 0.563 a_0$, and $k_c=0.46 a_0^{-1}$ for parameters in the first group, as well as the 
facet specific determinations, $V_R'= 7480$ meV, $V_R'= 5610$ meV, and $V_R'= 5210$ meV for Cu(111), Cu(100), 
and Cu(110), respectively.} 
in the modeling framework \cite{andersson1993,perandersson1993,persson2008}, Eqs.~(\ref{eq:1})-(\ref{eq:3}), to 
accurately reproduce the set of $\epsilon_n$ values.  The experiment-based PECs of Figure \ref{fig:1}(a) 
are characterized by minima position (separations from the last atom plane), and depths 
given as follows: 3.52 {\AA} and 29.0 meV for Cu(111), 3.26 {\AA} and 31.3 meV for Cu(100), 
2.97 {\AA} and 32.1 meV for Cu(110).  In Figure \ref{fig:1}(a), however, the curves are shown with minima 
positions slightly translated so that the set of $V_0(z)$ curves coincide at the classical turning 
point and thus facilitate an easy comparison.

The diffraction analysis of resonant back-scattering follows the
reasoning: For light adsorbates, like He and H$_2$, in gas-surface-scattering 
experiments, the elastic backscattering has a resonance structure. 
This provides a direct and elegant method to characterize the PEC, 
as they give accurate and detailed measurements of bound-level 
energies $\epsilon_n$ in the potential well. Isotopes with widely 
different masses, like ${}^3$He, ${}^4$He, H$_2$, D$_2$, 
permit a unique assignment of the levels and a determination of 
the well depth and ultimately a qualified test of model potentials~\cite{Roy}. 

For a resonance associated with a diffraction that involves a surface reciprocal lattice vector 
$\mathbf{G}$ there is a kinematical condition,
\begin{equation}
	\epsilon_i=\epsilon_n +\frac{\hbar^2}{2m_p}(\mathbf{K}_i-\mathbf{G})^2,
	\label{eq:5}
\end{equation}
where $m_p$ is the particle mass and where $\epsilon_i$ and 
$\mathbf{K}_i$ are the energy and wavevector component parallel to the 
surface of the incident beam, respectively. 
At resonance, weak periodic lateral corrugations of the basic 
interaction induce large changes in the diffracted beam intensities. 
The narrow resonance is observed as features in the diffracted beam 
intensities upon variations in the experimental incidence conditions. 
The intrinsically sharp resonances in angular and energy 
space have line widths that depend on intermediate bound-state life-time.
They are limited by elastic and phonon inelastic processes.
Lifetime broadening is only a fraction of a meV, substantially smaller 
than separations between the lower-lying levels (a few meV), allowing 
a number of physisorption levels $\epsilon_n$ with a unique assignment 
to be sharply determined from Eq.~(\ref{eq:5}).  

H$_2$
is the only molecule for which a detailed mapping of the bound-level
spectrum and the gas-surface interaction potential has been performed with
resonance scattering measurements
\cite{PerrauAndLapujoulade1982,YuEtAl1985,ChiesaEtAl1985,HartenEtAl1986,AnderssonEtAl1988,andersson1993,perandersson1993,persson2008}.
The sequences here were obtained using
nozzle beams of para-H$_2$ and normal-D$_2$, that is, the beams are
predominantly composed of $j = 0$ molecules.
Two isotopes H$_2$ and D$_2$ of widely different
masses and with the different rotational populations of
para-H$_2$  (p-H$_2$)
and ortho-D$_2$  (o-D$_2$) and the normal species (n-H$_2$, n-D$_2$)
are thus available; this richness in data means that the data analysis 
is greatly simplified and the interpretation is clear.
For instance, the rotational anisotropy of the interaction has been
determined via analysis of resonance structure resulting from the
rotational $(j, m)$  sub-level splittings observed for n-H$_2$ and
p-H$_2$ beams \cite{ChiesaEtAl1985,WilzenEtAl1991}.
Such knowledge permits a firm conclusion that the here-discussed measured
 bound-state energies, $\epsilon_n$ (Fig.~\ref{fig:2}),
refer to an isotropic distribution of
the molecular orientation.
The level assignment is compatible with a single
gas-surface potential for the two hydrogen isotopes~\cite{perandersson1993}.

Figure \ref{fig:2} illustrates the analysis that leads to a single accurate 
gas-surface potential curve for each of the facets from the experimentally 
observed energies $\epsilon_n$ \cite{andersson1993,perandersson1993,persson2008}.
The black open and filled circles represent measured $\epsilon_n$ values.
The procedure is an adaptation to surface physics of the Rydberg-Klein-Rees method of molecular physics 
\cite{Roy}. 
The  ordering is experimentally known and in this ordering, all $\epsilon_n$
values fall accurately on a common curve [when plotted
versus the mass-reduced level number $\eta = (n + 1/2)/\sqrt{m}$].
The variation in the quantization levels reflects the asymptotic behavior of the potential curve 
and thus determines the value of $C_\vdW$ to a high accuracy and gives
a good direct estimate of the well depth \cite{perandersson1993,persson2008}. 
A third-order polynomial fit to the data yields for $\eta = 0$ a
potential-well depth $D=29.5$, 31.4, and 32.3 meV for the (111), (100),
and (110) surfaces, respectively.
This direct construction of an effective
physisorption potential supplements the above-described experiment-based 
procedure [that instead uses the measured energies $\epsilon_n$ to fit 
$V_0(z)$ curves and obtain an even higher accuracy\footnote{We have tested 
the accuracy and consistency among the two experimental determinations 
of the effective physisorption potential for Cu(111). Specifically, we 
constructed an alternative $V_0(z)$ form in which we directly inserted 
the Rydberg-Klein-Rees/Le Roy value of $C_\vdW$ and used the directly 
extracted well depth 31.4 meV \cite{perandersson1993} to specify an 
effective value of $V_R'$.  We found that the lowest 4 physisorption 
eigenvalues of this potential then coincided within 3 percent of the 
measured $\epsilon_n$ energies. In contrast, the experiment-based 
potentials $V_0(z)$ described in Ref.\ \protect\cite{andersson1993} 
above and in potentials Fig.\ \protect\ref{fig:1}(a) reproduce the 
full set of measured energies (for all facets) to within 0.3 meV.}].

Figure \ref{fig:1}(a) also shows experiment-based determinations of the 
(rotationally averaged) amplitudes $V_1(z)$ of the lateral corrugation. 
The measured intensities of the first-order diffraction beams provide 
(as described above) an estimate of the resulting lateral variation
in the H$_2$-Cu potential. The corrugation is very small, 
$\sim$0.5 meV at the potential well minimum.
However, the existence of finite amplitudes $V_1(z)$ is essential:
The larger corrugation closer to the substrate contributes most 
importantly to the diffraction and resonance phenomena. In fact, it is 
a finite magnitude of $V_1(z)$ that ensures a coupling to the in-plane 
crystal momentum and allows an elastic scattering event to satisfy the 
kinematical condition (\ref{eq:5}).  

\section{van der Waals Accounts Used in DFT calculations}

Noncovalent forces, such as hydrogen bonding and vdW interactions, are 
crucial for the formation, stability, and function of molecules and 
materials. In sparse matter the vdW forces are particularly 
relevant in regions with low electron density. For a long time, it has 
been possible to account for vdW interactions  only by 
high-level quantum-chemical wave-function or Quantum Monte Carlo 
methods. The correct long-range interaction tail 
for separated molecules is absent from all popular local-density or 
gradient corrected XC functionals of density-functional theory, as well 
as from the Hartree-Fock approximation. Development of approximate 
DFT approaches that accurately model the 
London dispersion interactions \cite{Stone1997,Kaplan2006}  
is a very active field of research (reviewed in, for example 
Refs.~\onlinecite{GrimmeASM2007,GrafensteinCremer2009,JohnsonMD2009,SatoNakai2009,langreth2009}).
 
To account for vdW interactions in computational physics traditional 
DFT codes  are natural starting points. The vdW energy 
emanates from the correlated motion of electrons and there are 
proposals to account for it, like (i) DFT extended atom-pair potentials, (ii) explicit 
density functionals, and (iii) RPA in perturbation theory. Chemical 
accuracy is aimed for.  Extensive physical and chemical 
systems are of great interest, including bio- and nanosystems, where 
the vdW interactions are indispensable. Describing bonds in a variety of 
systems with ``chemical accuracy" requires that both strong and weak bonds 
are calculated. Strong covalent bonds are well described by traditional 
approximations, like the generalized gradient approximations (GGAs) 
\cite{PBE,LangrMehl,PerdewEtAl}, 
which are typically built in into approaches (i) and (ii). 
vdW-relevant systems range from small organic molecules to large and 
complex systems, like sparse materials and protein-DNA complexes. They 
all have noncovalent bonds of significance. Quite a number of naturally 
and technologically relevant materials have already been successfully  
treated \cite{langreth2009}.

The methods (i) and (ii) are essentially cost free, speed is given by 
that of traditional DFT (for example GGA based).
Such DFT calculations are competitive in terms of efficiency and broad 
applicability. Computational power is an important factor that is still 
relevant and an argument for choosing methods of types (i) and (ii) in many applications 
ahead of type (iii).

\subsection{DFT extended by atom-pair potentials} 

A common remedy for the missing vdW interaction in GGA-based DFT 
consists of adding a pairwise interatomic $C_6/R^6$ term ($E_\vdW$) 
to the DFT energy.
Examples are  
DFT-D \cite{Grimme2004}, TS-vdW \cite{TS}, and alike 
\cite{ElstnerHFSK2001,JureckaCHS2007}.
Refs.\ \cite{SunKLZ2008,ZhaoTruhlar2008} describe  
various other approaches also currently in use. 

The DFT-D method is a popular way to add on dispersion corrections to 
traditional Kohn-Sham (KS) density functional theory.  It is implemented into 
several code packages. Successively it has been refined to obtain higher 
accuracy, a broader range of applicability, and less empiricism. 
In the recent DFT-D3 version \cite{Grimme}, the main new ingredients 
are atom-pairwise specific dispersion coefficients and cutoff radii 
that are both computed from first principles. The coefficients for new 
eighth-order dispersion terms are computed using established recursion 
relations. Geometry-dependent information is here included by 
employing the new concept of fractional coordination numbers. 
They are used to interpolate between dispersion coefficients of atoms 
in different chemical environments. The method only requires adjustment 
of two global parameters for each density functional, is asymptotically 
exact for a gas of weakly interacting neutral atoms, and easily allows 
the computation of atomic forces. As recommended \cite{Grimme}, 
three-body nonadditivity terms are not considered. 

Another almost parameter-free\footnote{%
Both DFT-D and TS-vdW have a need to fix a cross-over function that is 
designed to minimize double counting of the semilocal correlation in regular
DFT and the vdW contribution. The parameter of this cross over is often 
fitted to a reference system, for example S22. 
}
 method for accounting for long-range vdW interactions 
from mean-field electronic structure calculations relies on the 
summation of interatomic $C_6/R^6$ terms, derived from the electron 
density of a molecule or solid and reference data for the free atoms \cite{TS}. 
The mean absolute error in the $C_6$ coefficients is 5.5\%, 
when compared to accurate experimental values for 1225 intermolecular pairs, 
irrespective of the employed exchange-correlation functional. 
The effective atomic $C_6$ coefficients have been shown to depend strongly 
on the bonding environment of an atom in a molecule \cite{TS}. 

\subsection{Explicit density functionals} 
Ground-state 
properties can be described by functionals of the electron density 
$n(\mathbf{r})$ \cite{Kohn}. 
The functional $E_{xc}[n(\mathbf{r})]$ for the XC energy is 
a central ingredient. The 
local-density approximation (LDA) \cite{KohnSham,HedinLu} and 
GGAs \cite{LangrMehl,PerdewEtAl,PBE}
do not describe the nonlocal correlations behind the vdW interactions. 
This subsection discusses explicit XC density functionals 
$E_{xc}[n(\mathbf{r})]$, focusing on the non-local correlation 
functional, $E_c^\nl[n(\mathbf{r})]$.

In the vdW-DF, the vdW interactions and correlations are 
expressed in terms of the density $n(\mathbf{r})$ as a truly nonlocal 
six-dimensional integral \cite{dion2004,langreth2005,thonhauser2007}. 
It originates in the adiabatic connection formula
\cite{PerdLangrI,GunnLund,PerdLangrII}, and uses an approximate 
coupling-constant integration and an approximate dielectric function 
with a single-pole form. 
The dielectric function is fully nonlocal and satisfies known limits, 
sum rules, and invariances, has a pole strength determined by a sum 
rule and is scaled to locally give the approximate gradient-corrected 
electron-gas ground-state energy. There are no empirical or fitted 
parameters, just references to general theoretical criteria.

Account for inhomogeneity is approximately
achieved by a gradient correction,
which is obtained from a relevant reference system.
In the original vdW-DF 
version~\cite{dion2004,thonhauser2007,rydberg2000,langreth2005}, 
the slowly varying electron gas is used for this. 
The gradient correction is then taken from Ref.~\onlinecite{LV1990}. 
Although promising results have been obtained for a variety of systems, 
including adsorption \cite{langreth2009,Mats}, there is room for improvements. 
Recently another reference system has been proposed, with the  
argument that adsorption systems have electrons 
in separate molecule-like regions, with exponentially decaying tails in 
between. The vdW-DF2 functional uses the gradient coefficient of the 
B88 exchange functional~\cite{B88} for the determination of the internal 
functional [Eq.~(12) of Ref.~\onlinecite{dion2004}] within the nonlocal 
correlation functional. This is based on application of the large-$N$ 
asymptote~\cite{Schwinger1980,Schwinger1981} on appropriate molecular 
systems. Using this method, Elliott and Burke~\cite{Elliott2009}  have 
shown, from first principles, that the correct exchange gradient coefficient 
$\beta$ for an isolated atom (monomer) is essentially identical to the 
B88 value, which had been previously determined empirically~\cite{B88}. 
Thus in the internal functional, vdW-DF2 \cite{LeeEtAl10} replaces 
$Z_{ab}$ in that equation with the value implied by the $\beta$ of B88. 
This procedure defines the relationship between the kernels of vdW-DF 
and vdW-DF2 for the nonlocal correlation energy. Like vdW-DF, vdW-DF2 
is a transferable functional based on physical principles and 
approximations. It has no empirical input.

The choices of exchange functional also differ. The original vdW-DF 
uses the revPBE~\cite{revPBE} exchange functional, which is good at 
separations in typical vdW complexes \cite{dion2004,langreth2005,thonhauser2007}. 
At smaller 
separations~\cite{puzder2006, kannemann-becke2009,murray2009,klimes2010,cooper2010}, 
recent studies suggest that the PW86 exchange functional \cite{PW86} 
most closely duplicates Hartree-Fock interaction energies both for atoms 
\cite{kannemann-becke2009} and molecules~\cite{murray2009}. 
The vdW-DF2 functional \cite{LeeEtAl10} employs the PW86R 
functional~\cite{murray2009}, which more closely reproduces the 
PW86 integral form at lower densities than those considered by the original PW86 authors.

\subsection{RPA} 
For first-principles electron-structure calculations, the random-phase 
approximation (RPA) to the correlation energy is presumably a 
suitable complement to the exact exchange energy \cite{HarlKresse2009}. 
The RPA to the correlation energy \cite{NozieresPines1958} 
incorporates a screened nonlocal exchange term and long-range dynamic 
correlation effects that underpin vdW bonding \cite{HarlKresse2009}. 

Hubbard, Pines and Nozieres \cite{pinesbook} pointed
out that RPA does have, at least, formal
limitations in the description of local correlations (large momentum
transfer). This is because RPA treats same- and opposite-spin scattering
on the same footing, thereby neglecting effects of Pauli exclusion in
the description of the RPA correlation term. There exist several
suggestions for RPA corrections \cite{mahan}. A recent
study \cite{Ren2011} suggests a single-excitation extension for RPA 
calculations
in inhomogeneous systems, thus lowering the mean average error for
noncovalent systems \cite{Ren2011}. 

Efficient RPA implementations
have become increasingly available for solids 
\cite{MiyakeGR2002,HarlKresse2009,MariniGR2006} and
molecular systems \cite{ScuseraHS,Furche2001,Furche2008,RenRS2009}.
One implementation \cite{HarlKresse2009} gives the XC functional as 
\begin{equation}
E_{xc} =  E_{\EXX} +  E_c,
\label{eq:6}
\end{equation}
where the exact exchange energy $E_{\EXX}$ (Hartree-Fock energy) and the 
correlation energy $E_c$, given as the independent-particle response 
function, are all evaluated from KS orbitals by using 
for example plane-wave code and suitably optimized projector augmented wave 
(PAW) potentials that describe high energy scattering properties very 
accurately up to 100 eV above the vacuum level \cite{ShishkinKresse2006}.  
The operations scale like $N^{6-7}$ and a high parallel efficiency can 
be reached \cite{HarlKresse2009}. 

\subsection{Implementation aspects\label{sec:3D}}
The vdW-DF2 calculations are performed by using the 
\textsc{abinit}~\cite{abinit1,abinit2} code with a plane-wave basis 
set and Troullier-Martins-type~\cite{Troullier1992} norm-conserving 
pseudopotentials. The scalar-relativistic correction is included in the 
pseudopotentials for transition-metals. A kinetic energy cut-off of 
70 Ry is used. For \emph{k}-space integrations, a $4\times{}4\times{}1$ 
Monkhorst-Pack mesh is used. For the partial occupation of metallic bands, 
we use the Fermi-Dirac smearing scheme with a 0.1~eV broadening width. 
With this setup the adsorption energies are converged within 1~meV. 
The vdW-DF total energy is calculated in a fully self-consistent way \cite{thonhauser2007}.
We adapted an implementation of the efficient vdW-DF algorithm~\cite{soler} 
from \textsc{siesta}~\cite{siesta} for use within a modified version of \textsc{abinit}.

\begin{figure}
\begin{center}
\includegraphics[width=0.3\textwidth]{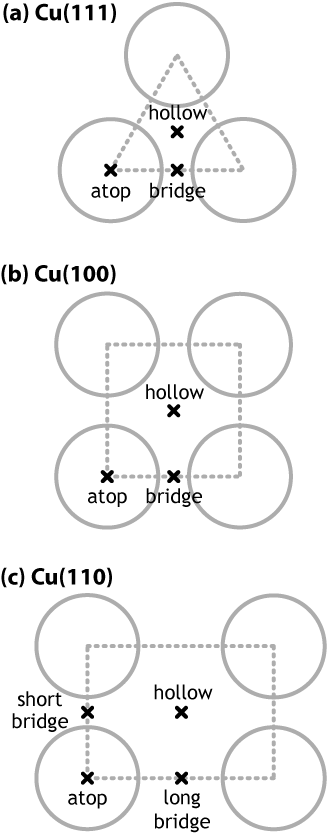}
\caption{\label{fig:4}
High-symmetry positions on the low-index Cu surfaces. In our calculations
the H$_2$ molecule lies flat in one of these positions.}
\end{center}
\end{figure}

The surfaces are modeled by a slab of four atomic layers with a vacuum
region of 20 \AA{} in a periodic supercell.  For the
calculations on the (111), (100), and (110) surfaces we use the surface
unit cells of $3\times2\sqrt{3}$, $3\times3$, and $2\sqrt{2}\times3$,
respectively.

In the electron-structure calculations, the molecule
is kept in a flat orientation above the high-symmetry
positions or sites\footnote{%
In the description of the DFT calculations we refer to these
positions as ``sites"
but note that there is a large zero-point
motion perpendicular to the surface and that the experimentally relevant
molecules move with a
large in-surface kinetic energy while trapped in the physisorption well.}
 on the Cu surfaces, as
indicated in Figure \ref{fig:4}.
Some test points indicate that the total energy depends very little on 
orientation. For instance, in a ``worst-case" situation, the energy 
change by a 90-degree in-plane rotation of H$_2$ at the long bridge 
site of Cu(110) is 0.92 meV, in a fixed-height in-plane rotation. 
If H$_2$ is moved to the equilibrium adsorption height, which is 
0.04 {\AA} lower, the energy difference increases to 0.97 meV. This variation is much smaller 
than the lateral corrugation ($\sim 4$ meV 
in this facet) and out-of-plane rotation ($\sim 5$ meV). 

To estimate the magnitude of the error introduced by neglecting the 
angular average (that automatically is included in the experimental data) we perform 
a separate calculation of H$_2$ in an up-right position on the Cu(111) 
surface using the optimal H$_2$-to-surface separation. This calculation 
results in a downward energy shift of 4.8 meV, which is the amplitude 
of the angular variation at the optimal separation. However, for an 
isotropic H$_2$ wave function this shift corresponds roughly to 
lowering of the ground state energy by merely $1/3 \times 4.8$ meV = 1.6 meV, 
assuming a simple sinusoidal energy variation in the angular space. 
The energy of higher order states would also be lowered, but to a 
lesser extent. This estimate thus indicates that the effect of the 
angular energy dependence is merely a minor quantitative correction 
to the eigenvalues. 

The PECs are calculated with vdW-DF2 for the high-symmetry sites, 
and from these the laterally averaged potential $V_0$ is approximately 
obtained. In turn, the bound quantum states in the $V_0$ potential 
well are calculated by solving the corresponding Schr\"odinger equation.

The theoretical values are generated for a number of discrete points, 
surface positions
(atop, hollow, long bridge, and short bridge), and 
separations $z$, while experimental data (Fig.\ \ref{fig:1}) are  
presented as lateral averages and functions of separation $z$. 
To connect the two, some approximation has to be made to 
extract the laterally averaged result out of the discrete one. 

To capture the effects of the surface topology and to use the 
surface-lattice points used in the DFT calculations, we find the 
following approximate average reasonable: 
\begin{equation}
V_0 = \frac{1}{4} ( V_\hollow+V_\myatop+V_\bridgea+V_\bridgeb)
\label{eq:7}		
\end{equation}
where $V_\bridgea=V_\bridgeb$ for the (100) and (111) surfaces. One argument 
in favor of this approximation is the fact that the (100) surface has 
twice as many bridge sites as there are atop sites. Approximation 
(\ref{eq:7}) is used in our comparisons between our vdW-DF2-determined 
potential $V_0$ and the  
experimentally determined one (Fig.\ \ref{fig:1}) and in our generation 
of eigenvalues used to relate to the experimental ones [which are 
defined in terms of a laterally averaged and H$_2$ rotational-angle 
averaged potential $V_0(z)$] (Fig.\ \ref{fig:2}). 
Reference \onlinecite{h2cu} uses the atop PEC on Cu(111), which is an adequate choice
due to the small corrugation of the Cu(111) surface.

For the discussion of the relation between corrugation and $V_1$, the 
classical turning point is the relevant separation. 
The corrugation of the PEC minimum is smaller, but still representative 
of the expected variation in the probability for the H$_2$ trapping.

The original report of experimental results also covers a potential 
$V_2(z)$. It represents the min-to-max variation of the lateral average 
of the rotational anisotropy. A full appreciation of the accuracy of the 
comparison between the experimentally determined $V_0(z)$ and  
$V_0^{\vdWDFtwo}(z)$, based on 
existing vdW-DF2 calculations, would benefit from an understanding of 
$V_2(z)$. To indicate that this is unlikely to have any large consequence 
for the three Cu surfaces, a simple estimate of the rotation angle effect 
is made above. This should be considered as a stimulus for further 
refinement of the testing.


\section{Results and Benchmarking}

We present a new benchmark, taken from surface physics,  with
extraordinary virtues.
Data are provided for
(i) energy eigenvalues, $\epsilon_n$, for H$_2$ and D$_2$ in the PEC well,
which have direct ties to measured reflection intensity,
(ii) the laterally averaged physisorption potential, $V_0$,
which is derived from measured data, the extracted PEC, and
(iii) the corrugation, $V_1$, also derived from measured data.

\subsection{Benchmarking strategies}

Evaluation of XC functionals is often made by comparing with other
theoretical results in a systematic way, for instance, the common
comparison with S22 data set
\cite{Jurecka2006,Sherrill2009, Molnar2009,Takatani2010,Szalewicz2010}.
These sets have twenty-two prototypical small molecular duplexes for
non-covalent interactions (hydrogen-bonded, dispersion-dominated, and mixed)
in biological molecules. They provide PECs at a very accurate level of
wave-function methods, in particular the CCSD(T) method. However, by
necessity, the electron systems in such sets are finite in size.
The original vdW-DF performs well on the S22 data set, except for
hydrogen-bonded duplexes (underbinding by about 15\% \cite{langreth2009,LeeEtAl10}).
Use of the vdW-DF2 functional reduces the mean absolute deviations of
binding energy and equilibrium separation significantly \cite{LeeEtAl10}.
Shapes of PECs away from the equilibrium separation are greatly improved.
The long-range part of the vdW interaction, particularly crucial for
extended systems, has a weaker attraction in the vdW-DF2, thus reducing
the error to 8 meV at separations 1~{\AA} away from equilibrium~\cite{LeeEtAl10}.

Recently, other numbers for the S22 benchmark on vdW-DF2 have been
published \cite{bligard}. The two calculations differ in the treatment
of the intermolecular separation, being relaxed \cite{LeeEtAl10} and
unrelaxed \cite{bligard}, respectively. Of course, absence of 
relaxations does lead to an appearance of worse performance.

Experimental information provides the ultimate basis for assessing
functionals. The vdW-DF functional has been promising in applications
to a variety of systems \cite{langreth2009}, but primarily vdW-bonded
ones. Typically, the calculated results are tested on measured
binding-energy and/or bond-length values that happen to be available.
The vdW-DF2 functional has also been successfully applied to some
extended systems, like graphene and graphite \cite{LeeEtAl10},
metal-organic-frameworks systems \cite{LeeEtAl10B}, molecular crystal
systems \cite{MolcrysDF2}, physisorption systems
\cite{LeeEtAl10C,selfassembly}, liquid water \cite{mogelhoj} and
layered oxides \cite{londero}. However, the studies are of the common
kind that focus on comparison against just a few accessible observations.

\begin{figure*}
\begin{center}
\includegraphics[width=0.95\textwidth]{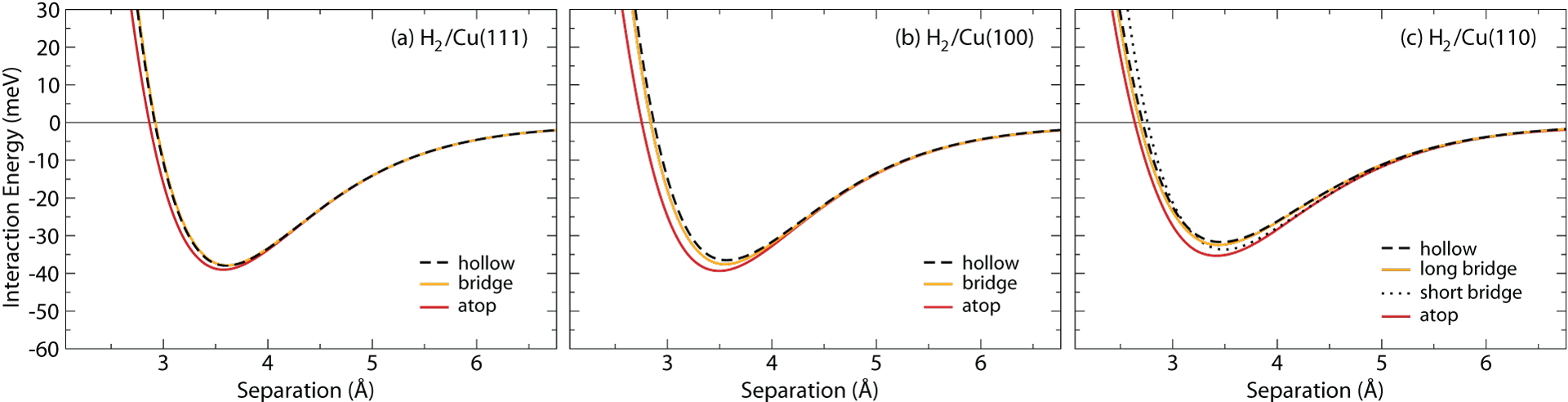}
\caption{\label{fig:5}
Calculated PECs for H$_2$ in atop, bridge, and hollow sites on the
(a) Cu(111), (b) (100), and (c) (110) surfaces,
calculated with the vdW-DF2 functional.}
\end{center}
\end{figure*}

Accurate
experimental values for the eigenenergies of H$_2$ and D$_2$ molecules
bound to Cu surfaces \cite{andersson1993,perandersson1993} motivate 
theoretical account and assessment. This knowledge base 
covers results for the whole shape of the physisorption potentials.
Here calculations on several Cu facets allow studies
of trends and a deeper analysis. The extensive report of vdW-DF2 results in
Figure \ref{fig:5} serves as a starting point.

\subsection{PECs from vdW-DF2}

PECs are calculated for H$_2$ in atop, bridge, 
and hollow sites on the Cu(111), (100), and (110) surfaces. 
The resulting benchmark for vdW-DF2 is also compared (below) with those 
of two other vdW approximations, 
the DFT-D3~\cite{Grimme} and TS-vdW \cite{TS} methods. 

To emphasize various aspects of the PECs and make valuable use of the 
numerical accuracy, the next few sections (and figures) highlight 
various aspects of the vdW-DF2 results.
For each position $z$ of the molecular center of mass,
we compare the averaged potential
functions $V_0(z)$ and $V_1(z)$, in Fig.\ \ref{fig:1},
and find strong qualitative agreement.
For instance, at the potential minimum of $V_0(z)$, Figure \ref{fig:1}
gives for $V_1(z)$ the approximate values 1, 3, and 4 meV for (111), (100),
and (110), respectively, in quantitative agreement with the illustration
of theoretical results analysis.
The insensitivity to facets and the corrugation is discussed in greater detail below.

\subsection{Isotropy of lateral averaged potentials}

An important 
feature of the experimental $V_0(z)$ in Figure \ref{fig:1} is the similar 
sizes of the well depths (range 29--32 meV) and 
separations (around 3.5 {\AA}) on the (111), (100), and 
(110) surfaces. This isotropy, i.e., similarity of physisorption 
potential among different facets, is interesting and perhaps surprising 
because Cu(111) contains a metallic surface state, whereas Cu(100) and 
Cu(110) do not. 

The most striking feature of the vdW-DF2 results for the H$_2$-Cu
PEC is probably that it is able to reproduce this isotropy.
From Figure \ref{fig:5} we find
physisorption depths in 
the interval 35--39 meV and separations in the range 3.3--3.6 {\AA}.
{}From the experimentally more relevant laterally averaged $V^\vdWDFtwo(z)$,
Fig.\ \ref{fig:1}(b), we find physisorption depths 31--36 meV.

\begin{figure}
\begin{center}
\includegraphics[width=0.4\textwidth]{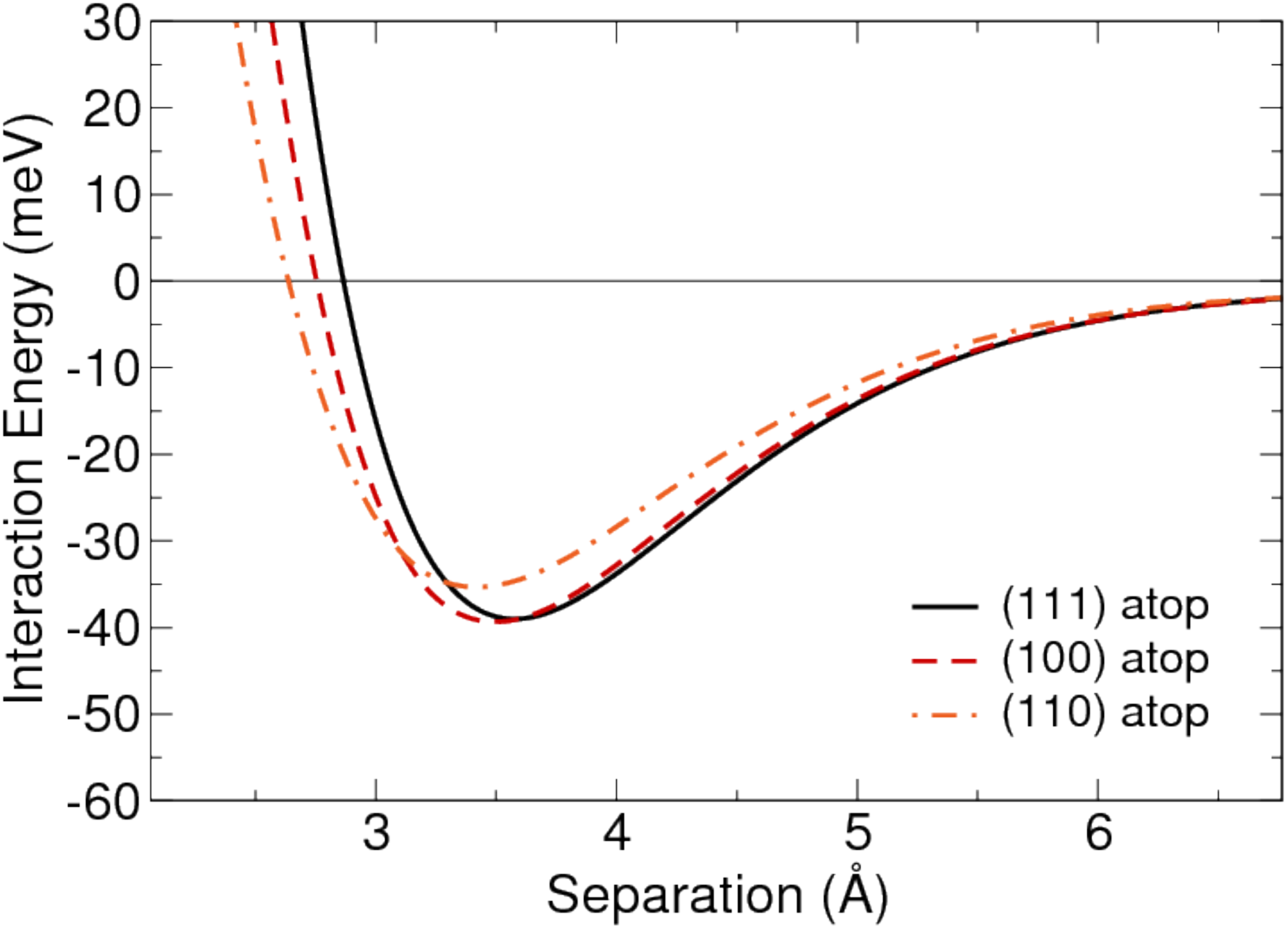}
\caption{\label{fig:6}
PECs of H$_2$ on atop sites of the Cu(111), (100), and (110) surfaces
calculated with the vdW-DF2 functional.}
\end{center}
\end{figure}

The isotropy is emphasized in Figure \ref{fig:6} by plotting the vdW-DF2 PECs on the atop 
sites of each surface. The curves lie very close to each other, both in 
the Pauli-repulsion region at short separations, dominated by $V_R$ in 
the traditional theory [Eqs.\ (\ref{eq:1})-(\ref{eq:3})], and in the 
vdW-attraction region at large separations. They differ only discernibly  
in the $V_R$-region, which can be understood in terms of the higher 
electron density, $n_o(\mathbf{r})$ on the atop site on the dense (111) 
surface.

Both agreements and differences are found in the trends (with facets) of
the experiment-based
[$V_0(z)$, $V_1(z)$] and vdW-DF2 based [$V^\vdWDFtwo(z)$] characterizations, 
Figure \ref{fig:1}(a) and (b). vdW-DF2
reproduces the trend in
ordering and roughly the magnitudes in the modulation amplitudes
[$V_1(z)$] at the classical
turning points (identified as position `0' in Figure \ref{fig:1}). As shown in
Fig.~\ref{fig:6}, vdW-DF2 also
reproduces 
the ordering of separations corresponding to physisorption minima, 
correctly decreasing as
(111) $>$ (100) $>$ (110). 
On the other hand, for $V_0(z)$ the physisorption depth
varies as  
 (111) $>$ (100) $>$ (110) whereas in 
$V^\vdWDFtwo(z)$ the depth varies (110) $>$ (100) $>$ (111). 
The largest relative difference, 25\%, is found for Cu(111). 
The set of physisorption depths on the three facets are reproduced
with an average confidence of 15\%.

\subsection{Corrugation\label{ssec:D}}

Another striking feature of the PECs is the variation with the density 
corrugation of each surface. In Figure \ref{fig:3}, the density profiles 
of the clean Cu(111), (100), and (110) surfaces indicate how the corrugation 
may vary. For fcc metals the (111) surface is the most dense, while the (100) 
and (110) surfaces are successively more open and thus corrugated. 
The trend is reflected in the PECs. These clear effects on the PEC are 
illustrated by the calculated PECs for H$_2$ in atop, bridge, and hollow 
sites on the Cu(111), (100), and (110) surfaces (Fig.\ \ref{fig:5}). 
{}From being small on the flat and dense (111) surface [Fig.\ \ref{fig:5}(a)], 
the corrugation grows from (111) to (100) and from (100) to (110), just 
as expected from the above reasoning.
 
\begin{figure}
\begin{center}
\includegraphics[width=0.4\textwidth]{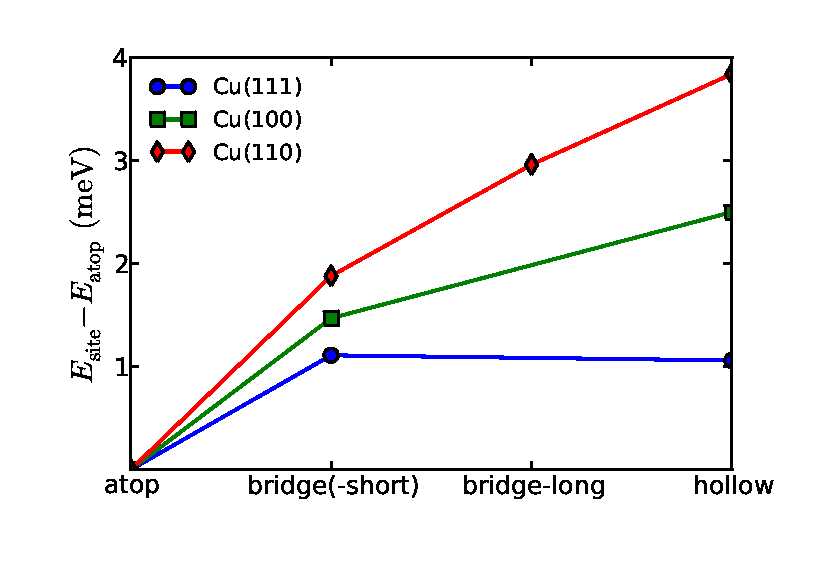}
\includegraphics[width=0.4\textwidth]{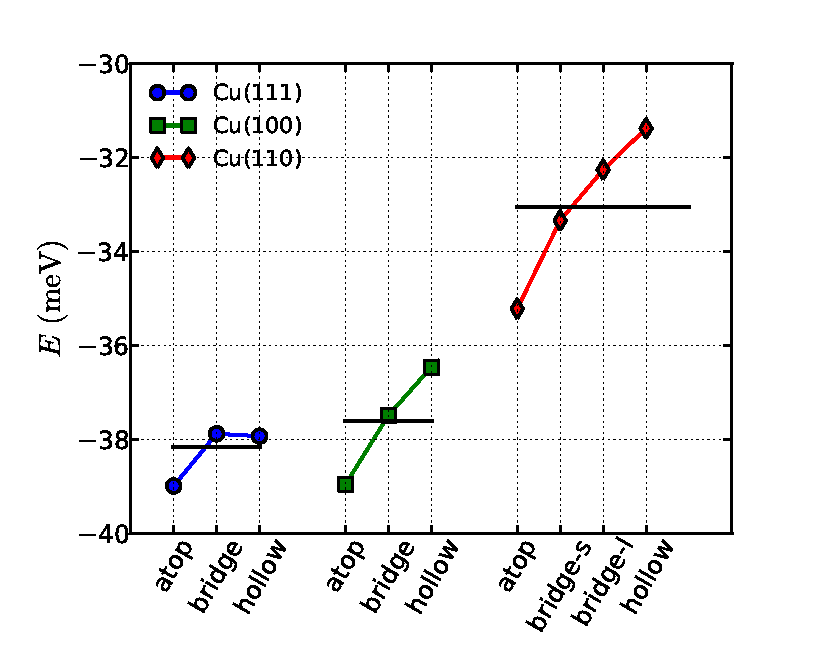}
\caption{\label{fig:7}
Corrugation for H$_2$ on Cu(111), Cu(100), and Cu(110) 
(a) illustrated by the lateral variation of the calculated adsorption
energy of H$_2$ at each high-symmetry position on these surfaces;
(b)
illustrated by calculated adsorption-energy values of H$_2$ on these
surfaces. The black horisontal lines indicate approximate site averages,
analogous to Eq.\ (\protect\ref{eq:7}). Subtracting the atop value from
each average gives 1.3, 1.6, and 2.5 meV as lateral variations on the
(111), (100) and (110) surfaces, these numbers can be interpreted as
measures of the corrugation.}
\end{center}
\end{figure}

Figure \ref{fig:7} shows the variations in the  adsorption                    
energies relative to the value of the atop configuration on the various facets.
We choose to report the corrugation at the PEC minimum.
The corrugation at the classical turning point is likely a stronger 
indicator of the strength of the elastic scattering that 
traps the incoming H$_2$ molecules, Sec.\ \ref{sec:2}.

On all three facets, the calculated stable site is atop 
(Figs.\ \ref{fig:5} and \ref{fig:7}). This result can be understood 
with  a simple 
argument based on the traditional model and a tight-binding (TB) 
description of the electrons. Equations (\ref{eq:1})--(\ref{eq:3}) 
separate the potential energy into repulsive and attractive parts, 
$V_R$  and $V_\vdW$, respectively. Close to the minimum point, the vdW 
attraction, $V_\vdW$ (Eq.\ (\ref{eq:3})) gets stronger in the direction 
towards the surface. The repulsion terms $V_R$ prevent the admolecule from 
benefiting from this by going even closer. Equation (\ref{eq:4}) reflects that 
higher density gives higher repulsion, and the density profiles in 
Figure \ref{fig:3} show the proper order. 
For a more general discussion, see Ref.\ \cite{Chen}.

While the electron density (Fig.\ \ref{fig:3}) is characterized by only 
one kind of corrugation, the corrugation of the PECs depends on where 
the probe hits the PEC. From Figure \ref{fig:5} we can envisage different 
corrugation values for different $z$ values. For the 
reflection-diffraction experiment one can argue that H$_2$ molecules 
coming in to the surface with a positive kinetic energy, i.e., at or 
above the energy of the classical turning point, are particularly relevant. 

\subsection{Corrugation and exchange functional} 

\begin{figure}
\begin{center}
\includegraphics[width=0.4\textwidth]{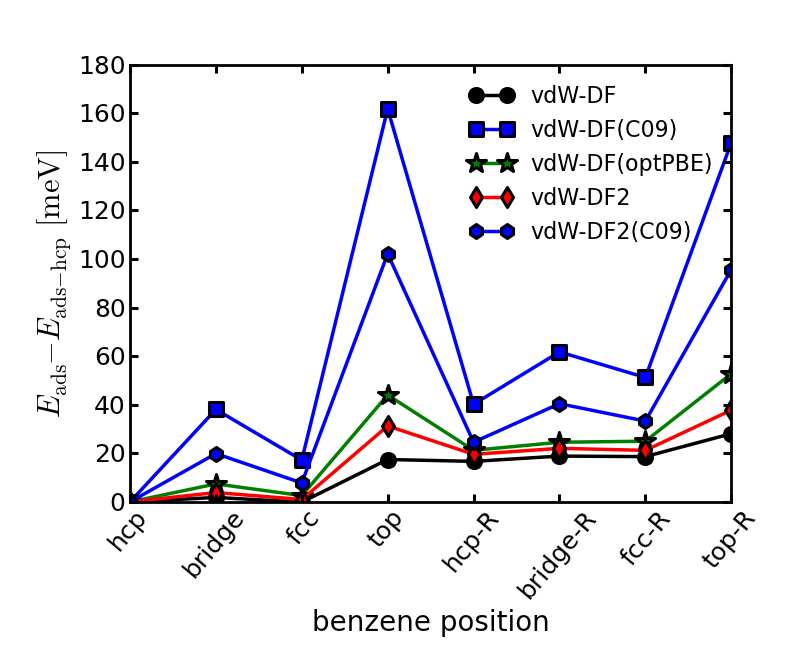}
\caption{\label{fig:8}
Interaction-energy values for benzene at various positions on the Cu(111)
surface,
calculated with different XC functionals. It shows that corrugation
energies are sensitive to choice of functional, according to calculations
with vdW-DF \cite{dion2004}, vdW-DF(C09) \cite{cooper2010},
vdW-DF(optPBE) \cite{klimes2010}, vdW-DF2 \cite{LeeEtAl10}, and vdW-DF2(C09).}
\end{center}
\end{figure}

To clarify the underlying cause of corrugation, a different adsorption system 
is first studied. Benzene on the Cu(111) surface is known to be a true 
vdW system \cite{slidingrings}. Interaction-energy values for the benzene 
molecule at various positions on the Cu(111) are calculated with five 
different density functionals, all accounting for vdW forces, and 
shown in Figure \ref{fig:8}. The functionals differ by the differing 
strengths of vdW attraction in vdW-DF and vdW-DF2, but in particular 
by different exchange approximations, 
revPBE \cite{dion2004,revPBE}, C09 \cite{cooper2010}, 
optPBE \cite{klimes2010}, and PW86R \cite{PW86,murray2009,LeeEtAl10}. 
While the binding-energy value is not so sensitive to choice of functional, 
corrugation-energies values are. 

Therefore, the accurate results of a reflection-diffraction experiment are 
valuable in several respects. On one hand, they support that the 
traditional model is right in its separation in 
Eqs.\ (\ref{eq:1})-(\ref{eq:3}) and there attaching the repulsive wall, 
$V_R$, to Pauli repulsion \cite{persson2008} and thus to exchange. 
On another hand, they are able to discriminate between different 
approximations for the exchange functional. Similar results have also 
been calculated for benzene on graphene \cite{berland12}.	

\subsection{Comparison with experiment-related quantities}

\begin{figure}
\begin{center}
\includegraphics[width=0.4\textwidth]{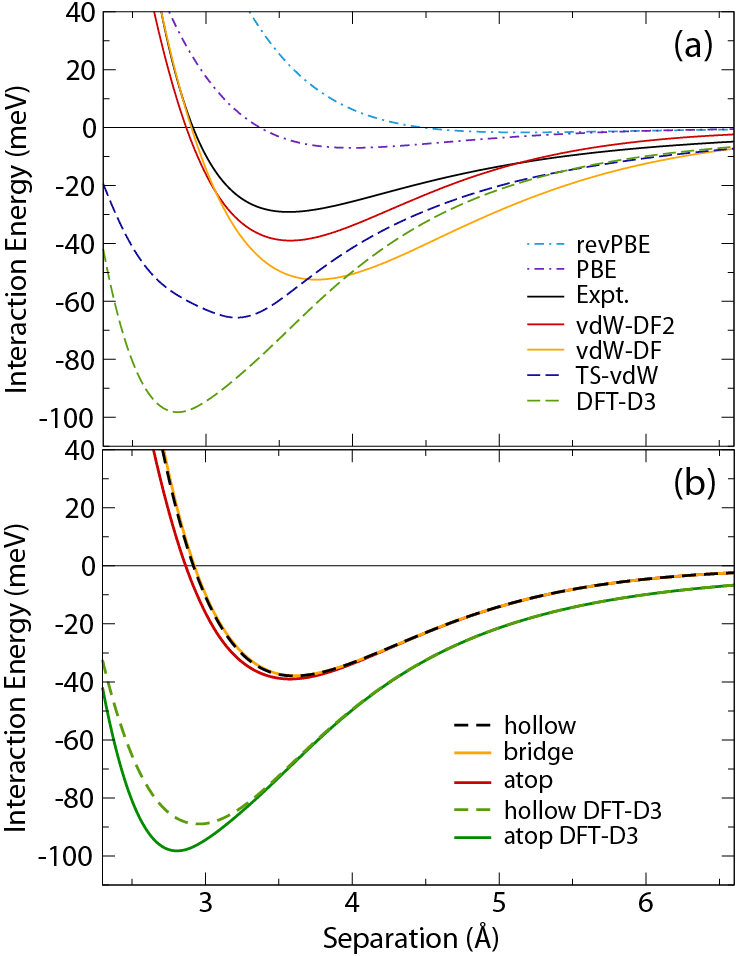}
\caption{\label{fig:9}
(a) Experimentally determined effective physisorption potential for H$_2$
at atop site on the Cu(111) surface~\protect\cite{andersson1993},
compared with potential-energy curves for H$_2$ on the Cu(111) surface,
calculated for the atop site in GGA-revPBE, GGA-PBE, vdW-DF2,
vdW-DF \cite{h2cu}, TS-vdW \cite{TS}, and DFT-D3(PBE) \cite{Grimme}.
Partly adapted from Ref.\ \onlinecite{h2cu}.
(b) Comparison of PECs for atop and hollow sites on Cu(111) calculated
with vdW-DF2 and DFT-D3.
}
\end{center}
\end{figure}

Comparison of the calculated results on the H$_2$-Cu systems with the 
model used for the analysis of the experimental 
data \cite{andersson1993, perandersson1993} is next done for the 
laterally averaged potential $V_0(z)$, the potential derived from 
experiment. The experiment-derived results in Figure \ref{fig:1} are 
redrawn in Figure \ref{fig:9} as the experimental physisorption 
potential for H$_2$ on Cu(111). Figure \ref{fig:9}(a) shows our 
comparison of PECs calculated using vdW-DF and vdW-DF2, respectively, 
drawn against the experimental physisorption potential for H$_2$, 
originally published in Ref.\ \onlinecite{h2cu}. The Cu(111) surface 
is chosen for its flatness that gives clarity in the analysis and 
eliminates several side-issues that could have made interpretations fuzzier. 
Several qualitative similarities are found for both vdW-DF and vdW-DF2 functionals. 
The vdW-DF2 functional gives PECs in a useful qualitative and 
quantitative agreement with the experimental physisorption curve, for 
instance with respect to well depth, equilibrium separation, and 
curvature of PEC near the well bottom, and thus zero-point vibration frequency. 
Comparisons of full PECs are also parts of the benchmarking. 

\subsection{Other methods}

Well depths and equilibrium separations for H$_2$ on Cu(111) are seen 
as PEC minimum points in Figure \ref{fig:9}(a). As the corrugation is 
so small, it should suffice to use only the atop result. More precise 
value pairs \cite{h2cu} are 
($-28.9$ meV; 3.5 {\AA}) for 
$V_0(z)$ extracted from experiment \cite{andersson1993}, 
($-53$ meV; 3.8 {\AA}) as calculated with the vdW-DF functional, 
($-39$ meV; 3.6 {\AA}) with the vdW-DF2 functional, 
($-98$ meV; 2.8 {\AA}) with the DFT-D3(PBE) method \cite{Grimme}, and 
($-66$ meV; 3.2 {\AA}) with the TS-vdW method \cite{TS}. 
The striking discrepancy between the three major types of accounts for 
vdW in extended media is discussed below. 

As reported for example in Ref.\ \onlinecite{h2cu}, the LDA and GGA 
functionals do not describe the nonlocal correlation effects that give 
vdW forces. They also misrepresent the PECs. The minima are too shallow 
and the equilibrium separations are too large \cite{rydberg2000}. 

In Figure \ref{fig:9}, the DFT-D3(PBE) curves are the results of calculations 
with the DFT-D3 corrections \cite{Grimme} added on top of the PBE PECs. 
Figure \ref{fig:9}(b) compares a DFT-D3 PEC at the hollow site of the (111) 
surface with that of an atop site [same as the one in Fig.\ \ref{fig:9}(a)] 
\cite{h2cu}. The energy difference between atop and hollow 
adsorption-energy values is found to be 11 meV. If this corrugation at 
$z_\mymin$ were a measure of the corrugation, the corrugation by DFT-D3 
on Cu(111) would be 11 meV, or 11 times larger than that given by vdW-DF2 (1 meV).

It is interesting to note that the TS-vdW method delivers a PEC 
[Fig.\ \ref{fig:9}(a)] similar to that from DFT-D3 
[Fig.\ \ref{fig:9}(a) and (b)], although not quite as deep.

\subsection{Energy levels}

The quantum-mechanical motion of the H$_2$ molecule in the various 
(laterally averaged) potentials (Fig.\ \ref{fig:1}) can be calculated 
and for the motion perpendicular to the surface be described by, e.g., 
the bound-state eigenenergy values. 
Figure \ref{fig:2} presents and compares results from experiment and theory.
The experimental curve, identified by filled (H$_2$) and empty (D$_2$) 
black circles may be analyzed 
\cite{andersson1993,perandersson1993,persson2008} within the traditional 
theoretical picture \cite{zaremba1977,HarrisNordlander} of the interaction 
between inert adsorbates and metal surfaces, Section \ref{sec:2}.
The experimental level sequence in Figure \ref{fig:2} can be accurately reproduced 
($<0.3$ meV) by such a physisorption potential 
\cite{perandersson1993,persson2008} (Fig.\ \ref{fig:1}), having a well 
depth of 28.9 meV and a potential minimum located 3.5~{\AA} outside 
the topmost layer of copper ion cores. From the measured intensities of 
the first-order diffraction beams, a very small lateral variation of 
the H$_2$-Cu(111) potential can be deduced, $\sim 0.5$ meV at the 
potential-well minimum.

The vdW-DF2 theory results for the levels are identified by filled 
(H$_2$) and empty (D$_2$) red circles. The theoretical results 
are constructed from the calculated vdW-DF2 PECs in Figure \ref{fig:5} 
by first providing an estimate for the laterally averaged potential 
$V^{\vdWDFtwo}_0(z)$ for each facet, according to 
Eq.\ (\ref{eq:7}). 

We note that unlike the experimental results [which define 
$V_0(z)$], the variation in $V^{\vdWDFtwo}_0(z)$ does not 
reflect an average over the angles of the H$_2$. 
We also note that this is a small effect, Section \ref{sec:3D}.

Figure \ref{fig:2} documents good agreement in results from vdW-DF2 
and experiment for the energy levels on each of the Cu facets. 
The eigenvalues have the same order as in the experimental results 
(Fig.\ \ref{fig:2}), indicating 
good agreement between the calculated 
and measured average potentials.

There are some discrepancies between the eigenvalues for 
H$_2$. This signals that the vdW-DF2 functional might not give the right 
shape for the PEC of H$_2$ on Cu(111). The vdW-DF2 PEC is judged to lie 
close to the experimental physisorption potential, both at the equilibrium 
position and at separations further away from the surface, and is thus 
described as ``promising" \cite{h2cu}. The same applies for H$_2$ on 
Cu(100) and Cu(110).

\subsection{Summary}

Having access to the full PEC, including shape of potential and 
asymptotic behavior, allows a more stringent assessment of the 
theoretical results. This is in addition to the many other 
conclusions that Figure \ref{fig:9} gives. 

\textit{PECs from vdW-DF2:\/} 
The picture for H$_2$ on Cu(111) of Ref.\ \onlinecite{h2cu} at large 
applies also to the Cu(100) and (110) surfaces. 
Going from the most dense 
surface, Cu(111), to the more open surfaces, the changes are small. 
The vdW-DF2  description is good to 25\% in calculating the potential depth
 in the worst case, Cu(111). 
vdW-DF2 describes the mean physisorption well depths (averaged over
all three facets) to within 15\% of the experiments.

\textit{Lateral average:\/}
Approximate lateral averages of the PECs, $V_0$, have a fair agreement 
with those derived from experiment (Fig.\ \ref{fig:1}).

\textit{Isotropy:\/} On the fcc metal Cu, the PECs of H$_2$ are almost 
isotropic (Figs.\ \ref{fig:5} and \ref{fig:6}).
 
\textit{Corrugation:\/} 
On each surface, the PECs vary (Figs.\ \ref{fig:5}--\ref{fig:8}) with 
the density corrugation (Fig.\ \ref{fig:3}). For fcc metals the (111) 
surface is most dense, and (100) and (110) are successively more open 
and corrugated.  For the calculated PECs for H$_2$ in atop, bridge, 
and hollow sites on the Cu, trends and magnitudes of $V_1$ agree with 
experimental findings (Figs.\ \ref{fig:1} and \ref{fig:7}).

\textit{Corrugation and exchange functional:\/} 
By the example of the benzene molecule on the Cu(111) surface, 
calculations with several different density functionals that account 
for vdW forces show that corrugation-energies values are sensitive to 
functionals that differ by different exchange approximations. 
Therefore, the accurate results of a reflection-diffraction experiment 
are valuable for discriminating between exchange functionals 
(Fig.\ \ref{fig:8}). The vdW-DF2 functional uses a good exchange approximation.

\textit{Comparison with experiment-related quantities:\/} 
The experiment-derived results are shown in Figures \ref{fig:1} and 
\ref{fig:9} as the experimental physisorption potential for H$_2$ on Cu(111). 
Comparison of PECs calculated with vdW-DF and vdW-DF2 show that the 
vdW-DF2 functional gives PECs in a useful qualitative and quantitative 
agreement with the experimental physisorption curve.

\textit{Other functionals:\/} 
PECs calculated with several different methods for H$_2$ on Cu(111) 
in atop and hollow positions
show a striking discrepancy between the 
results from the DFT-D3 \cite{Grimme} and TS-vdW \cite{TS} methods
on the one hand, 
and those of vdW-DF2 and experiment on the other hand. 
This discrepancy is traced back to the fact that pair potentials
center the interactions on the nuclei
and do not fully reflect
that important binding contributions arise in the wave function tails
outside the surface.

\textit{Energy levels:\/} 
The energy levels in the H$_2$-Cu PEC wells (Fig.\ \ref{fig:2}) 
are calculated for all facets and compared with the experimental ones 
(Fig.\ \ref{fig:2}). Agreement with experimental results is gratifying.

We judge the performance of vdW-DF2 as very promising.
In making this assessment we observe that
(i) vdW-DF2 is a first-principles method, where 
characteristic electron energies are typically in the eV range, and 
(ii) the test system and results are very demanding.
The second point is made evident by the fact that
other popular 
methods deviate significantly more from the experimental curve. 
For instance, application of the DFT-D3(PBE) method \cite{Grimme} 
(with atom-pairwise specific dispersion coefficients and cutoff radii 
computed from first principles) gives ($-98$ meV; 2.8 {\AA}) for the 
PEC minimum point. 
The good agreement of the minima of the vdW-DF2 and experimental curves 
are encouraging, and so is the relative closeness of experimental 
and calculated eigenenergy values in Figure \ref{fig:2}. 
The discrepancies between the eigenvalues signal that the vdW-DF2 PEC 
might not 
express the exact shape of the physisorption potential
for H$_2$ on Cu(111).

\section{Comparisons and Analysis}
	
Figures \ref{fig:9}(a) and (b) show that the vdW-DF2 functional 
and DFT-D3 and TS-vdW methods give very different results. We could 
have made this point even stronger by showing also results for corrugation 
and energy values.
However, we believe that comparisons of the PECs at atop and hollow sites 
on the Cu(111) surface suffice. No doubt, H$_2$ on Cu is a demanding 
case for all methods. The local probe H$_2$ on Cu avoids smearing-out 
effects, unlike for example graphene and PAHs, and incipient covalency, 
unlike H$_2$O and CO, and is thus a pure vdW system and a lateral-sensitive one.

We trace the differing of the results with the vdW-DF2 and DFT-D3 methods to 
the differences in the descriptions of the vdW forces. The vdW-DF2 and 
similar functionals describe interactions between all electrons, while 
DFT-D3 and TS-vdW methods rely on atom-pair interactions. Even if large 
efforts are put into mimicking the real electron-charge distribution by 
electron-charge clouds around each atom nucleus, this has to be a 
misrepresentation of surface-induced redistributions of electronic 
charge in a general-geometry correction of a traditional DFT calculation. 
There is no mechanism for Zaremba-Kohn effects. 

The Zaremba-Kohn 
formulation of physisorption is not build in explicitly into the 
vdW-DF2 functional. However, the interactions of the electrons are 
build in into the supporting formalism in the similar way as in the 
derivation of the Zaremba-Kohn formula. 

\section{Conclusions}

Accurate and extensive experimental data are used to benchmark 
calculational schemes for sparse matter, that is, methods that account 
for vdW forces. Reflection-diffraction experiments on light particles 
on well-characterized surfaces provide accurate data banks of experimental 
physisorption information, which challenge any such scheme to produce 
relevant physisorption PECs. PECs of H$_2$ on the Cu(111), (100), and 
(110) surfaces are here studied. Accuracy is high even by the
surface-physics standards and is here provided thanks to diffraction 
kinematical conditions giving sharp resonances in diffraction beam intensities.
We propose that such surface-related PEC benchmarking should find a broader usage. 

The vdW-DF, vdW-DF2, DFT-D3, and TS-vdW schemes are used, and results 
are compared.  The first two are expressions of the vdW-DF method, that is, 
nonempirical nonlocal functionals in which the electrodynamical couplings
of the plasmon response produces fully distributed contributions to vdW interactions;
Like in the Zaremba-Kohn picture \cite{zaremba1977}, they permit the extended 
conduction electrons to respond also in the density tails outside a surface.
The latter two are examples of DFT extended with vdW pair potentials and represent 
the dispersive interaction through an effective response and pair potentials located on 
the nuclei positions.  Several qualitative similarities are found between the vdW-DF 
and vdW-DF2 functionals.  The vdW-DF2 functional gives PECs in a useful qualitative and 
quantitative agreement with the experimental PECs. This is looked at for well depths, 
equilibrium separations, and curvatures of PEC near the well bottom, and thus molecular
zero-point vibration frequency. The DFT-D3 and TS-vdW schemes give PEC results that 
deviate significantly more from experimental PECs. The benchmark with the experimental H$_2$/Cu 
scattering data is thus able to discriminate between the results of pair-potential-extended DFT 
methods and vdW-DF2. The differences suggest that it is important to 
reflect the actual, distributed location of the fluctuations (plasmons) 
that give rise to vdW forces. 

The vdW-DF2 density functional benchmarks very well against the S22 data 
sets \cite{LeeEtAl10}. 
It is also the functional being more extensively compared between 
experiment and theory here. 
Certain very well-pronounced features, like isotropy of the H$_2$-Cu 
PEC, the (111), (100), and (110) PECs being close to identical, and the 
clear trend in its corrugation that grows in order 
(111) $<$ (100) $<$ (110), are well described. 
The calculated  $V_1$ results are found to be close to the experimental
ones, thereby being almost decisive on exchange functionals.
The energy levels for the quantum-mechanical 
motion in the H$_2$-Cu PEC agree in a gratifying way.

The accuracy of this experiment is also shown to be valuable for discriminating between exchange 
functionals (Fig.\ \ref{fig:6}). The vdW-DF2 functional is found to apply a good exchange approximation.

The vdW-DF2 is found promising for applications at short and intermediate 
distances, as is relevant for adsorption. However, the accuracy of 
experimental data is high enough to stimulate a more detailed 
analysis of all aspects of the theoretical description. This should be 
valuable for the further XC-functional development. For instance, some 
discrepancies are found for the eigenvalues. They signal that the 
vdW-DF2 PEC for H$_2$ on Cu(111) might not have a perfect shape. 
Additional physical effects could be searched for. The metallic surface 
state on Cu(111) might be one source; It is possible that the metallic 
nature of the H$_2$/Cu(111), H$_2$/Cu(100), and H$_2$/Cu(110) systems 
motivates modifications in the description of the electrodynamical 
response inside the nonlocal functional. For a well established 
conclusion, a more accurate theory is called for.
 
In any case, H$_2$/Cu physisorption constitutes possibilities for 
benchmarking theory descriptions and
represents a very strong challenge for the density functional development. 

\acknowledgments
The Swedish National Infrastructure for Computing (SNIC) at C3SE is 
acknowledged for providing computer allocation and the Swedish Research 
Council (VR) for providing support to KB, ES, and PH. 
Work by KL is supported by NSF DMR-0801343, MY is sponsored by the
US Department of Energy, Basic Energy Sciences, Materials Sciences and Engineering Division. 



\begin{thebibliography}{99}

\bibitem{Kohn} 
P. Hohenberg and W. Kohn, Phys. Rev. \textbf{136}, B864 (1964). 

\bibitem{KohnSham} 
W. Kohn and L.J. Sham, Phys. Rev. \textbf{140}, A1133 (1965). 

\bibitem{LeeEtAl10}
K. Lee, \'E.D. Murray, L. Kong, B.I. Lundqvist, and
D.C. Langreth, Phys. Rev. B (RC) \textbf{82}, 081101 (2010).

\bibitem{Grimme}
S. Grimme, J. Antony, S. Ehrlich, and H. Krieg,
J. Chem. Phys. \textbf{132}, 154104 (2010).

\bibitem{TS}
A. Tkatchenko and M. Scheffler,
Phys. Rev. Lett. \textbf{102}, 073005 (2009).

\bibitem{dion2004}
M. Dion, H. Rydberg, E. Schr\"oder,
D.C. Langreth, and B.I. Lundqvist, Phys. Rev. Lett.
\textbf{92}, 246401 (2004); \textbf{95}, 109902(E) (2005).

\bibitem{HarlKresse2009}
J. Harl and G. Kresse,
Phys. Rev. Lett. \textbf{103}, 056401 (2009).

\bibitem{Jurecka2006}
P. Jure\v{c}ka, J. \v{S}poner, J. \v{C}ern\'y, and P. Hobza,
Phys. Chem. Chem. Phys. \textbf{8}, 1985 (2006).

\bibitem{Takatani2010}
T. Takatani, E.G. Hohenstein, M. Malagoli, M.S. Marshall,
and C.D. Sherrill, J. Chem. Phys. \textbf{132}, 144104 (2010).

\bibitem{Svetla}
S. D. Chakarova-K\"ack, E. Schr\"oder, B. I. Lundqvist, and D. C. Langreth,
Phys. Rev. Lett. \textbf{96}, 146107 (2006).

\bibitem{andersson1993} 
S. Andersson and M. Persson, Phys. Rev. Lett. \textbf{70}, 202 (1993). 

\bibitem{perandersson1993} 
S. Andersson and M. Persson, Phys. Rev. B  \textbf{48}, 5685 (1993). 

\bibitem{persson2008} 
See, e.g., M. Persson and S. Andersson, Chapter 4, ``Physi\-sorption Dynamics at Metal Surfaces", in 
Handbook of Surface Science, Vol. 3 (Eds. E. Hasselbrink and B.I. Lundqvist), Elsevier, 
Amsterdam (2008), p. 95. 

\bibitem{h2cu}
K. Lee, A.K. Kelkkanen, K. Berland, S. Andersson, D.C. Langreth,
E. Schr\"oder, B.I. Lundqvist, and P. Hyldgaard,
Phys. Rev. B \textbf{84}, 193408 (2011).

\bibitem{Roy}
R.J. Le Roy, Surf. Sci. \textbf{59}, 541 (1976).

\bibitem{SchGunnars1980}
K. Sch\"onhammer and O. Gunnarsson,
Phys. Rev. B \textbf{22}, 1629 (1980).

\bibitem{Brenig1987}
W. Brenig, 
Physica Scripta \textbf{35}, 329 (1987).

\bibitem{Boato}
G. Boato, P. Cantini, and R. Tatarek, in
\textit{Proceedings of the Seventh International Vacuum Congress
and the Third International Conference on Solid Surfaces},
Vienna, 1977, edited by R. Dobrozemsky et al. (F. Berger and Sohne,
Vienna, 1977), p. 1377.

\bibitem{Garcia}
N. Garcia, J. Ibaniz, J. Solana, and N. Canbrera,
Surf. Sci. \textbf{60}, 385 (1976).

\bibitem{zaremba1977}
E. Zaremba and W. Kohn,
Phys. Rev. B \textbf{15}, 1769 (1977).

\bibitem{HarrisNordlander} 
P. Nordlander and J. Harris,
J. Phys. C \textbf{17}, 1141 (1984).

\bibitem{jelliumedge}
N.D. Lang,
Solid State Physics \textbf{28}, 225 (1973).

\bibitem{ZarembaKohn1976}
E. Zaremba and W Kohn,
Phys. Rev. B \textbf{13}, 2270 (1976).

\bibitem{Liebsch1986}
A. Liebsch, 
Europhys. Lett. \textbf{1}, 361 (1986).

\bibitem{Esbjerg}
N. Esbjerg and J. K. N{\o}rskov,
Phys. Rev. Lett. \textbf{45}, 807 (1980).

\bibitem{Smoluchowski1941}
R. Smoluchowski, Phys. Rev. \textbf{60}, 661 (1941).

\bibitem{EMT}
K.W. Jacobsen, J.K. N{\o}rskov, and M.J. Puska,
Phys. Rev. B \textbf{35}, 7423 (1987).

\bibitem{PerrauAndLapujoulade1982}
J. Perrau and J. Lapujoulade,
Surf. Sci. \textbf{121}, 341 (1982).

\bibitem{YuEtAl1985}
C.-F. Yu, K.B. Whaley, C. Hogg, and S. Sibener,
J. Chem. Phys. \textbf{83}, 4217 (1985).

\bibitem{ChiesaEtAl1985}
M. Chiesa, L. Mattera, R. Musenich, and C. Salvo,
Surf. Sci. \textbf{151}, L145 (1985).

\bibitem{HartenEtAl1986}
U. Harten, J.P. Toennies, and C. W\"oll,
J. Chem. Phys. \textbf{85}, 2249 (1986).

\bibitem{AnderssonEtAl1988}
S. Andersson, L. Wilzen, and M. Persson,
Phys. Rev. B \textbf{38}, 2967 (1988).

\bibitem{WilzenEtAl1991}
L. Wilzen, F. Althoff, S. Andersson, et al.,
Phys. Rev. B \textbf{43}, 7003 (1991).

\bibitem{HarrisAndLiebsch1982b}
J. Harris and A. Liebsch,
J. Phys. C - Solid State Physics \textbf{15}, 2275 (1982).

\bibitem{Stone1997}
A.J. Stone,
The Theory of Intermolecular Forces,
Oxford University Press, Oxford, 1997.

\bibitem{Kaplan2006}
G. Kaplan,
Intermolecular Interactions,
Wiley, Chichester, 2006.

\bibitem{GrimmeASM2007}
S. Grimme, J. Antony, T. Schwabe, and
C. M\"uck-Lichtenfeld, Org. Biomol. Chem. \textbf{5}, 741 (2007).

\bibitem{GrafensteinCremer2009}
J. Gr\"afenstein and D. Cremer,
J. Chem. Phys. \textbf{130}, 124105 (2009).

\bibitem{JohnsonMD2009}
E.R. Johnson, I.D. Mackie, and G.A. DiLabio,
J. Phys. Org. Chem. \textbf{22}, 1127 (2009).

\bibitem{SatoNakai2009}
T. Sato and H. Nakai, 
J. Chem. Phys. \textbf{131}, 224104 (2009).

\bibitem{langreth2009}
D.C. Langreth, B.I. Lundqvist,
S.D. Chakarova-K\"ack, V.R. Cooper, M. Dion, P. Hyldgaard,
A. Kelkkanen, J. Kleis, L. Kong, S. Li, P.G. Moses,
E. Murray, A. Puzder, H. Rydberg, E. Schr\"oder, and T.
Thonhauser, J. Phys.: Cond. Mat. \textbf{21}, 084203 (2009).

\bibitem{PBE}
J.P. Perdew, K. Burke, and M. Ernzerhof,
Phys. Rev. Lett. \textbf{77}, 3865 (1996); \textbf{78}, 1396(E) (1997).

\bibitem{LangrMehl}
D.C. Langreth and M.J. Mehl,
Phys. Rev. Lett. \textbf{47}, 446 (1981).

\bibitem{PerdewEtAl}
J. P. Perdew et al.,
Phys. Rev. Lett. \textbf{100}, 136406 (2008).

\bibitem{Grimme2004}
S. Grimme, J. Comput. Chem. \textbf{25}, 1463 (2004).

\bibitem{ElstnerHFSK2001}
M. Elstner, P. Hobza, T. Frauenheim, S. Suhai, and E. Kaxiras,
J. Chem. Phys. \textbf{114}, 5149 (2001).

\bibitem{JureckaCHS2007}
P. Jure\v{c}ka, J. \v{C}ern\'y, P. Hobza, and D.R. Salahub,
J. Comput. Chem. \textbf{28}, 555 (2007).

\bibitem{SunKLZ2008}
Y. Y. Sun, Y.-H. Kim, K. Lee, and S. B. Zhang,
J. Chem. Phys. \textbf{129}, 154102 (2008).

\bibitem{ZhaoTruhlar2008}
Y. Zhao and D.G. Truhlar,
Acc. Chem. Res. \textbf{41}, 157 (2008).

\bibitem{HedinLu}
L. Hedin and B.I. Lundqvist,
J. Physics Part C Solid State Physics \textbf{4}, 2064 (1971). 

\bibitem{thonhauser2007}
T. Thonhauser, V.R. Cooper,
S. Li, A. Puzder, P. Hyldgaard, and D.C. Langreth,
Phys. Rev. B \textbf{76}, 125112 (2007).

\bibitem{langreth2005}
D.C. Langreth, M. Dion, H. Rydberg, E. Schr\"oder, P. Hyldgaard, and 
B.I. Lundqvist, 
Int. J. Quant. Chem. \textbf{101}, 599 (2005).

\bibitem{PerdLangrI}
D.C. Langreth and J.P. Perdew,
Solid State Commun. \textbf{17}, 1425 (1975).

\bibitem{GunnLund}
O. Gunnarsson and B.I. Lundqvist,
Phys. Rev. B \textbf{13}, 4274 (1976).

\bibitem{PerdLangrII}
D.C. Langreth and J.P. Perdew,
Phys. Rev. B \textbf{15}, 2884 (1977).

\bibitem{rydberg2000}
H. Rydberg, B.I. Lundqvist, D.C. Langreth, and M. Dion,
Phys. Rev. B \textbf{62}, 6997 (2000).

\bibitem{LV1990}
D.C. Langreth and S.H. Vosko, in
``Density Functional Theory of Many-Fermion Systems",
ed. S.B. Trickey, Academic Press, Orlando, 1990.

\bibitem{Mats}
Y.N. Zhang, F. Hanke, V. Bortolani, M. Persson, and R. Q. Wu,
Phys. Rev. Lett. \textbf{106}, 236103 (2011).

\bibitem{B88}
A.D. Becke, Phys. Rev. A \textbf{38}, 3098 (1988).

\bibitem{Schwinger1980}
J. Schwinger,
Phys. Rev. A \textbf{22}, 1827 (1980).

\bibitem{Schwinger1981}
J. Schwinger,
Phys. Rev. A \textbf{24}, 2353 (1981).

\bibitem{Elliott2009}
P. Elliott and K. Burke,
Can. J. Chem.  \textbf{87}, 1485 (2009).

\bibitem{revPBE}
Y. Zhang and W. Yang,
Phys. Rev. Lett. \textbf{80}, 890 (1998).

\bibitem{puzder2006}
A. Puzder, M. Dion, and  D.C. Langreth,
J. Chem. Phys. \textbf{126}, 164105 (2006).

\bibitem{kannemann-becke2009}
F.O. Kannemann and A.D. Becke,
J. Chem. Theory Comput. \textbf{5}, 719 (2009).

\bibitem{murray2009}
{\'E}.D. Murray, K. Lee, and D.C. Langreth,
Jour. Chem. Theor. Comput. \textbf{5}, 2754 (2009).

\bibitem{klimes2010}
J. Klime\v{s}, D.R. Bowler, and A. Michaelides,
 J. Phys.: Condens. Matter \textbf{22}, 022201 (2010).

\bibitem{cooper2010}
V.R. Cooper, 
Phys. Rev. B \textbf{81}, 161104(R) (2010).

\bibitem{PW86}
J.P. Perdew and Y. Wang,
Phys. Rev. B \textbf{33}, 8800(R) (1986).

\bibitem{NozieresPines1958}
P. Nozi\`eres and D. Pines,
Phys. Rev. \textbf{111}, 442 (1958).

\bibitem{pinesbook}
D. Pines and P. Nozi\`eres,
``The Theory of Quantum Liquids," Vol 1 (Addison-Wesley Publ. Comp.,
inc. Redwood City, 1966), p. 327.

\bibitem{mahan}  G.D. Mahan,
``Many-particle Physics" (2nd ed.), 
Plenum Press, New York (1990), pp. 444--454.

\bibitem{Ren2011}
X. Ren, A. Tkatchenko, P. Rinke, and M. Scheffler,
Phys. Rev. Lett. \textbf{106}, 153003 (2011).

\bibitem{MiyakeGR2002}
T. Miyake et al., Phys. Rev. B \textbf{66}, 245103 (2002).

\bibitem{MariniGR2006}
A. Marini, P. Garc\'ia-Gonz\'alez, and A. Rubio,
Phys. Rev. Lett. \textbf{96}, 136404 (2006).

\bibitem{ScuseraHS}
G.E. Scuseria, T.M. Henderson, and D.C. Sorensen,
J. Chem. Phys. \textbf{129}, 231101 (2008).

\bibitem{Furche2001}
F. Furche,
Phys. Rev. B \textbf{64}, 195120 (2001).

\bibitem{Furche2008}
F. Furche,
J. Chem. Phys. \textbf{129}, 114105 (2008).

\bibitem{RenRS2009}
X. Ren, P. Rinke, and M. Scheffler,
Phys. Rev. B \textbf{80}, 045402 (2009).

\bibitem{ShishkinKresse2006}
M. Shishkin and G. Kresse,
Phys. Rev. B \textbf{74}, 035101 (2006).

\bibitem{abinit1}
X. Gonze et al.,
Zeit. Kristallogr. \textbf{220}, 558 (2005).

\bibitem{abinit2}
X. Gonze et al.,
Computer Phys. Commun. \textbf{180}, 2582 (2009).

\bibitem{Troullier1992}
N. Troullier and J. L. Martins,
Phys. Rev. B \textbf{46}, 1754 (1992).

\bibitem{soler}
G. Rom\'an-P\'erez and J.M. Soler,
Phys. Rev. Lett.  \textbf{103}, 096102 (2009).

\bibitem{siesta}
P. Ordej\'on, E. Artacho, and J.M. Soler,
Phys. Rev. \textbf{53}, 10441(R) (1996);
J.M. Soler, E. Artacho, J.D. Gale, A. Garc\'ia, J. Junquera, P. Ordej\'on, and
D. S\'anchez-Portal,
J. Phys.: Condens. Matter \textbf{14}, 2745 (2002).

\bibitem{Sherrill2009} 
D. Sherrill, T. Takatani, and E. G. Hohenstein,
J. Phys. Chem. A \textbf{113}, 10146 (2009).

\bibitem{Molnar2009}
L.F. Molnar, X. He, B. Wang, and K.M. Merz,
J. Chem. Phys. \textbf{131}, 065102 (2009).

\bibitem{Szalewicz2010}
R. Podeszwa, K. Patkowski, and K. Szalewicz,
Phys. Chem. Chem. Phys. \textbf{12}, 5974 (2010).

\bibitem{bligard}
J. Wellendorff and T. Bligaard,
Topics in Catalysis \textbf{54}, 1143 (2011).

\bibitem{LeeEtAl10B}
L.Z. Kong, G. Rom\'an-P\'erez, J.M. Soler, and D.C. Langreth,
Phys. Rev. Lett. \textbf{103}, 096103 (2009).

\bibitem{MolcrysDF2}
K. Berland, {\O}. Borck, and P. Hyldgaard,
Comp. Phys. Commun. \textbf{182}, 1800 (2011).

\bibitem{LeeEtAl10C}
K. Lee, Y. Morikawa, and D.C. Langreth,
Phys. Rev. B \textbf{82}, 155461 (2010).

\bibitem{selfassembly}
J. Wyrick, D.-H. Kim, D. Sun, Z .Cheng, W. Lu, Y. Zhu, K. Berland,
Y.S. Kim, E. Rotenberg, M. Luo, P. Hyldgaard, T.L. Einstein, and
L. Bartels, Nano Letters \textbf{11}, 2944 (2011).

\bibitem{mogelhoj}
A. M{\o}gelh{\o}j, A. Kelkkanen, K.T. Wikfeldt,
J. Schi{\o}tz, J.J. Mortensen, L.G.M. Pettersson, B.I. Lundqvist,
K.W. Jacobsen, A. Nilsson, and J.K. N{\o}rskov,
J. Phys. Chem. B \textbf{115}, 14149 (2011).

\bibitem{londero}
E. Londero and E. Schr\"oder,
Computer Phys. Commun. \textbf{182}, 1805 (2011).

\bibitem{Chen}
De-Li Chen, W.A. Al-Saidi, and J.K. Johnson,
Phys. Rev. B \textbf{84}, 241405R (2011).

\bibitem{slidingrings}
K. Berland, T.L. Einstein, and P. Hyldgaard,
Phys. Rev. B \textbf{80}, 155431 (2009).

\bibitem{berland12}
K. Berland et al., unpublished (2012).

\end{thebibliography}
\end{document}